\documentclass[aps,prl,reprint]{revtex4-1}

\def\jcap{{J. Cosmology Astropart. Phys.}}
\def\prd{{Phys.~Rev.~D}}        
\usepackage{amsmath,amstext}
\usepackage[T1]{fontenc}
\usepackage{amssymb}
\input{epsf}
\usepackage{graphicx}
\usepackage{ae,aecompl}

\usepackage{hyperref}
\usepackage{amsmath}
\usepackage{amssymb}
\usepackage{mathtools}
\usepackage{bm}
\usepackage{cleveref}
\usepackage{tensor}
\usepackage{braket}
\usepackage{enumitem}
\usepackage{mhchem}

\usepackage{color}

\newcommand{\spc}{\quad \quad \quad}

\newcommand{\riga}{\textcolor{white}{-}}

\def\be{\begin{equation}}
\def\ee{\end{equation}}
\def\beq{\begin{eqnarray}}
\def\eeq{\end{eqnarray}}

\begin{document}
\title{Boosting unstable particles}
\author{L.~Gavassino$^1$, and F.~Giacosa$^{2,3}$}
\affiliation{$^1$Nicolaus Copernicus Astronomical Center, Polish Academy of Sciences, ul. Bartycka 18, 00-716 Warsaw, Poland}
\affiliation{$^2$Institute of Physics, Jan-Kochanowski Univ., ul. Uniwersytecka 7, 25-406 Kielce, Poland}
\affiliation{$^3$Institute for Theoretical Physics, J.  W. Goethe University, Max-von-Laue-Str. 1, 60438 Frankfurt, Germany.}

\begin{abstract}
In relativity, there is no absolute notion of simultaneity, because two clocks that are in different places can always be desynchronized by a Lorentz boost. Here, we explore the implications of this effect for the quantum theory of unstable particles. We show that, when a wavefunction is boosted, its tails travel one to the past and the other to the future. As a consequence, in the new frame of reference, the particle is in a quantum superposition ``decayed + non decayed'', where the property ``decayed-ness'' is entangled with the position. Since a particle cannot be localised in a region smaller than the Compton wavelength, there is a non-zero lower bound on this effect, which is fundamental in nature. The surprising implication is that, in a quantum world, decay probabilities can never be Lorentz-invariant. We show that this insight was the missing ingredient to reconcile the seemingly conflicting views about time dilation in relativistic quantum mechanics and quantum field theory. 
\end{abstract} 

\maketitle

\textit{Introduction -} The problem of how to rigorously formulate a relativistic quantum theory for unstable particles has been a subject of debate for sixty years \cite{Zwanziger1963,Kawai1969,Weldon1976,
Alicki1986,Stefanovich1996,
shirokov2004,shirokov2005,
stefanovich2006,stefanovich2006,Alavi:2014mxa,Urbanowski:2014gza,UrbaRazy2014,Urbanowski:2017haf,Exner:1983xu,ExnerCheck1973,Stefanovich:2005ai,Giacosa:2015mpm,Giacosa:2018dzm,Stefanovich:2018xji}. Although a lot of progress has been made, two fundamental questions still remain unanswered:
\begin{itemize}
\item Is it possible for two observers in relative motion to disagree on whether an unstable particle is in a decayed state or not \cite{Exner:1983xu,ExnerCheck1973,Stefanovich:2005ai,Giacosa:2015mpm}?
\item Concerning the decay law of moving particles, are there any quantum corrections to the relativistic dilation of time \cite{shirokov2004,shirokov2005,
stefanovich2006,stefanovich2006,Alavi:2014mxa,Urbanowski:2014gza}?
\end{itemize}
Clearly, these questions have very broad relevance, since in highly energetic events (such as supernovae, cosmic-ray showers, accelerator experiments, and the early Universe) unstable particles travel in space with very high speeds \cite{Bailey:1977de,CERN-Mainz-Daresbury:1978ccd,Baerwald2012,Lipari2012,
Farley2015,SCHRODER2017,Jaeckel2018}. The topic also has important implications for neutrino physics, as all constraints on neutrino lifetimes \cite{Joshipura2002,Gomes2015,Chack2021,Moss2018} have time dilation as an built-in assumption.

The goal of this article is to finally resolve the debate around the above questions, in a way that is both rigorous and intuitive. We will show that the seemingly contradictory results found by many authors \cite{Stefanovich1996,
shirokov2004,shirokov2005,
stefanovich2006,stefanovich2006,Alavi:2014mxa,Urbanowski:2014gza,UrbaRazy2014,Urbanowski:2017haf,Exner:1983xu,ExnerCheck1973,Stefanovich:2005ai,Giacosa:2015mpm,Giacosa:2018dzm,Stefanovich:2018xji} are a necessary consequence of the \textit{relativity of simultaneity} (the mechanism by which two clocks are desynchronized in a Lorentz boost \cite{Wald,special_in_gen,GavassinoSuperluminal2021}). In a nutshell, we will prove that, when a particle is unstable, position uncertainty is Lorentz-transformed into ``decayed-ness'' uncertainty, because the simultaneity hyperplane is redefined. As a consequence, the decay probability is \textit{not} a Lorentz scalar.


Throughout the paper, we adopt the signature $(-,+,+,+)$ and work in natural units $c=\hbar=1$.
For exposition purposes, we take the neutron, which is unstable to $\beta$-decay \footnote{Note that the lifetime of the neutron is subject to uncertainties due to incompatible results obtained with experimental  different methods \cite{Wietfeldt:2011suo}. It has been speculated that this anomaly is due to Beyon-Standard-Model physics \cite{Fornal:2018eol,Berezhiani:2018eds} or the Anti-Zeno effect \cite{Giacosa:2019nbz}.}
\begin{equation}\label{decayneutron}
n \rightarrow p^+ + e^- + \bar{\nu}_e \, ,
\end{equation}
as our reference particle. However, our results can be straightforwardly generalised to any unstable particle.

\textit{ The ``Alavi-Giunti argument'' -} 
It is useful, as a first step, to review a couple of apparently contradictory arguments, which are actually the key to understanding our article. The first argument is due to \citet{Alavi:2014mxa}. According to them, ``being a neutron'', or ``being a proton + an electron + a neutrino'', are absolute factual truths (valid in all reference frames), because neutrons and, e.g, protons have very different observational signatures. They reason that, if a neutron passes through a detector, it leaves a different track with respect to a proton, and such track can be seen by all observers, independently from their state of motion. 

Let us make this argument a little more formal, by considering a concrete observable. The electric four-current $j^{\mu}(x)$ transforms under a Lorentz boost $\Lambda$ as below \footnote{Equation \eqref{magnetic} is just the transformation law of a vector field \cite{MTW_book,Weinberg_book_1972} in Quantum Field Theory \cite{wightman_book,BjorkenDrell_book,
nakanishi_book,
Ticciati1999,Peskin_book}. For a rigorous proof of \eqref{magnetic} in the context of (fully interacting) quantum electrodynamics see appendix B of \citet{Zumino1960}. Note that equation \eqref{magnetic} is valid also in the context of (fully interacting) Relativistic Quantum Dynamics, see equation (9.4) of \citet{Keister1991}.}:
\begin{equation}\label{magnetic}
U^\dagger (\Lambda) j^{\mu}(x) U(\Lambda) = \Lambda\indices{^\mu _\rho} \, j^{\rho }(\Lambda^{-1} x) \, .
\end{equation}
Here, ${U}(\Lambda)$ is the unitary representation of $\Lambda$. Averaging \eqref{magnetic} over a state $\ket{\alpha}$, defining $\ket{\Lambda\alpha}:= {U}(\Lambda)\ket{\alpha}$, and setting $x=0$, we obtain
\begin{equation}\label{boostofaraday}
\bra{\Lambda \alpha}  j^{\mu}(0) \ket{\Lambda \alpha} = \Lambda\indices{^\mu _\rho}  \bra{\alpha} j^{\rho}(0) \ket{\alpha} \, .
\end{equation}
Now, it is evident that, if $\ket{\alpha}$ models an isolated neutron at rest near the origin, it will impress a characteristic ``neutronic footprint'' on $\bra{\alpha} j^{\rho} \ket{\alpha} $. In fact, a neutron does not have a net charge, but it carries a measurable Amp\`{e}rian magnetic moment (i.e., a closed of loop of electric current \cite{Jackson1977,Boyer1988,Mezei1988,ParticleDataGroup:2020ssz}). On the other hand, equation \eqref{magnetic} tells us that, when we make a Lorentz boost, the quantum average of the electric four-current transforms like a classical vector. Hence, the boost sets the magnetic moment in motion. But this implies that we cannot interpret the state $\ket{\Lambda\alpha}$ as $p^+ + e^- + \bar{\nu}_e$, because two sharply separated charges (the proton and the electron) cannot be confused with a single (connected \footnote{All observers agree on the spacetime topology \cite{Hawking1973} of the support of a tensor field.}) loop of electric four-current. Thus, a neutron in proximity of the origin is ``perceived as a neutron'' by all observers who sit in the origin, independently from their state of motion.

\textit{The ``Exner-Stefanovich theorem'' -} There is a simple mathematical theorem \cite{Exner:1983xu,ExnerCheck1973,Stefanovich:2005ai} that seems to contradict the reasoning above. Let's take a look at it. Suppose that there is a projector $\mathcal{Q}$, which returns ``$\, 1 \,$'' if the state models a neutron, and ``$ \, 0 \, $'' otherwise. If $K_1$ is the generator of the boosts in the direction $1$, and $P^1$ is the first component of the four-momentum, we can write down the Jacobi identity:
\begin{equation}\label{Jacobu}
[{\mathcal{Q}},[{K}_1,{P}^1]]+[{K}_1,[{P}^1,{\mathcal{Q}}]]+[{P}^1,[{\mathcal{Q}},{K}_1]]=0 \, .
\end{equation}
On the other hand, $[{K}_1,{P}^1]=i{H}$, where $H$ is the Hamiltonian \cite{weinbergQFT_1995}. Furthermore, if a state $\ket{\alpha}$ has a certain probability of being a neutron, the state $e^{-iP^j a_j}\ket{\alpha}$, which is just a copy of $\ket{\alpha}$ translated in space, should have exactly the same probability of being a neutron. Hence, $\mathcal{Q}$ is invariant under space translations:
\begin{equation}\label{traslocomesenoncifosse}
e^{iP^j a_j} \, \mathcal{Q} \, e^{-iP^j a_j} = \mathcal{Q} \spc (\forall \, a_j \in \mathbb{R}^3) \, .
\end{equation}
This implies that $[P^j,\mathcal{Q}]=0$, and equation \eqref{Jacobu} becomes 
\begin{equation}\label{hqpqk}
i[{H},{\mathcal{Q}}] = [{P}^1,[{\mathcal{Q}},{K}_1]] \, .
\end{equation}
Since the neutron decays, the operator ${\mathcal{Q}}$ cannot be a conserved quantity. Therefore, $[{H},{\mathcal{Q}}] \neq 0$. It follows from equation \eqref{hqpqk} that also $[\mathcal{Q},K_1]\neq 0$, which implies
\begin{equation}\label{projector}
U^\dagger (\Lambda) \, \mathcal{Q} \, U(\Lambda) \neq \mathcal{Q} \, .
\end{equation} 
This is telling us that, if $\ket{\alpha}$ is a neutron, it is not guaranteed that also $\ket{\Lambda \alpha}$ will be a neutron. This seems to be in stark contrast with the argument of \citet{Alavi:2014mxa}. But, is there really a contradiction? 

\begin{figure}
\begin{center}
\includegraphics[width=0.5\textwidth]{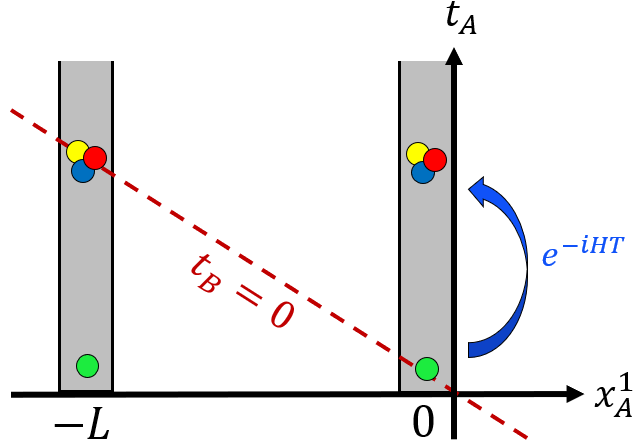}
	\caption{Minkowski diagram of our thought experiment. In Alice's frame, there is $\{1/2,1/2\}$ probability of having a neutron (green circles) in either of two boxes (grey areas), at $t_A=0$. After a time $T$, the neutron decays into $p^++e^-+\bar{\nu}_e$ (blue+yellow+red circles) in both boxes. Bob moves with velocity $-v$ with respect to Alice. His line of contemporary events (red dashed line) is oblique, and intersects the two boxes at two different times for Alice. As a consequence, in Bob's frame there is $1/2$ probability of having a neutron in the right box, and $1/2$ probability of having a proton, an electron, and a neutrino in the left box. Therefore, the Lorentz boost has entangled two observables: the neutron projector $\mathcal{Q}$, and the position of the center of mass of the system.
	}
	\label{fig:boxes}
	\end{center}
\end{figure}

\textit{A thought experiment -} Consider the following experimental set up. In Alice's frame, $\{t_A,x_A^j\}$, there are two small boxes at rest, which are kept closed. One box is located at $x_A^1=0$. The other box is located at $x_A^1=-L$, where $L>0$ is a very large distance. At $t_A=0$, a neutron $n$ is in a pure state $\ket{\psi}$, with $1/2$ probability of being in one box, and $1/2$ probability of being in the other box:
\begin{equation}\label{sblurf}
  \ket{\psi}= \dfrac{\ket{n\text{ in box ``} {-L}  \text{''}} + \ket{n\text{ in box ``} 0  \text{''}}}{\sqrt{2}} \, .  
\end{equation}
After some time (say, $T=\text{``}5\text{ lifetimes of }n\text{''}$), the neutron is transformed, by unitary evolution, into $p^++e^-+\bar{\nu}_e$, inside both boxes, with probability $1-e^{-5}\approx 1$. Hence,
\begin{equation}
 e^{-i{H}T} \! \ket{\psi} \approx  \dfrac{\ket{p,e,\bar{\nu}\text{ in box ``} {-L}  \text{''}} + \ket{p,e,\bar{\nu}\text{ in box ``} 0 \text{''}}}{\sqrt{2}} \, .
\end{equation}
The Minkowski diagram of this process is shown in figure \ref{fig:boxes}.
Now, suppose that Bob moves with velocity $-v$ with respect to Alice, and assume that $vL \equiv T$. What is the state of the neutron in Bob's frame at $t_B=0$, assuming that Bob is in the origin? The hyperplane $\{ t_B=0\}$ coincides with the hyperplane $\{ t_A=-vx_A^1 \}$, and is plotted in figure \ref{fig:boxes}. As we can see, it intersects the two boxes at two different Alice's times. In particular, the left box intersects the hyperplane in the event $(T,-L)$, while the right box intersects the hyperplane in the origin. On the other hand, we know that at $(T,-L)$ the neutron has decayed, while in the origin it has not decayed yet. Therefore, if $\Lambda$ is the boost that connects Alice and Bob, namely
\begin{equation}
\Lambda = 
\begin{bmatrix}
\gamma & \gamma v \\
\gamma v & \gamma \\
\end{bmatrix}  ,
\end{equation}
we can write
\begin{equation}\label{lambdasblurs}
U(\Lambda)\ket{\psi} \approx  \dfrac{\ket{p,e,\bar{\nu}\text{ in box ``} {-L}  \text{''}} + \ket{n\text{ in box ``} 0 \text{''}}}{\sqrt{2}} \, .
\end{equation}
Recalling the definition of $\mathcal{Q}$, we immediately see that
\begin{equation}
1 = \bra{\psi}\mathcal{Q}\ket{\psi} \neq  \bra{\psi}U^\dagger (\Lambda) \, \mathcal{Q} \, U(\Lambda)\ket{\psi} \approx \dfrac{1}{2} \, .
\end{equation}
The physical meaning of equation \eqref{projector} is finally clarified: in the relativistic transformation of time, $t_B=\gamma(t_A+vx_A^1)$, the term ``$\, vx_A^1 \, $'' can convert future events into present events, anticipating a decay. This effect becomes stronger the further the particle is from the origin. As a consequence, in equation \eqref{lambdasblurs}, the ``decayed-ness'' is correlated with the position. Measuring which box is heavier (i.e. where the particles are) automatically collapses the wavefunction into a state in which the neutron has decayed with a probability that is either $0$ or $1$.

Note that the present thought experiment does not contradict the argument of \citet{Alavi:2014mxa}: if two observers look at the same spacetime \textit{event}, they agree on whether such event contains a neutron or its decay products (because a loop of four-current cannot be Lorentz-transformed into two point charges). On the other hand, by relativity of simultaneity, two observers can disagree on whether that specific event belongs to the past, present, or future. This is the physical mechanism by which a boost can effectively ``cause a decay''.

\textit{A more formal proof -} For completeness, we provide here a more formal derivation of equation \eqref{lambdasblurs}. Suppose that $\ket{\alpha}$ models a neutron at rest in the origin. Then, $\bra{\alpha}\mathcal{Q}\ket{\alpha}=1$. Since the origin is a fixed point of Lorentz boosts ($\Lambda 0 =0$), we can invoke the argument of \citet{Alavi:2014mxa}, and assume that $\ket{\Lambda\alpha}= {U}(\Lambda)\ket{\alpha}$ is still a neutron: $\bra{\Lambda\alpha}\mathcal{Q}\ket{\Lambda\alpha} \approx 1$. Now, let's consider the state
\begin{equation}
\ket{\Lambda \beta} := U(\Lambda)\ket{\beta} \, , \quad \text{with} \quad  \ket{\beta}:=  e^{iP^1 L} \ket{\alpha} \, .
\end{equation}
Using the transformation law of the four-momentum \cite{srednicki_book},
\begin{equation}\label{poincorro}
U(\Lambda) 
\begin{bmatrix}
 H \\
 P^1 \\
\end{bmatrix}
U^\dagger (\Lambda) = \Lambda^{-1}
\begin{bmatrix}
 H \\
 P^1 \\
\end{bmatrix}
=\gamma
\begin{bmatrix}
H-v P^1 \\
P^1-vH \\
\end{bmatrix} \, ,
\end{equation}
we can rewrite $\ket{\Lambda\beta}$ as follows:
\begin{equation}
\ket{\Lambda\beta}=U(\Lambda)e^{iP^1 L} \ket{\alpha}=  e^{i P^1 \gamma L} e^{-i H \gamma v L} \ket{\Lambda\alpha}  \, .
\end{equation}
Averaging $\mathcal{Q}$ over $\ket{\Lambda\beta}$, and recalling equation \eqref{traslocomesenoncifosse}, we obtain
\begin{equation}\label{gvL}
\bra{\Lambda\beta} \mathcal{Q}\ket{\Lambda\beta} =\bra{\Lambda \alpha} e^{i H \gamma v L} \mathcal{Q} e^{-i H \gamma v L} \ket{\Lambda\alpha} \xrightarrow{ L \to \infty } 0 \, .
\end{equation}
As we can see, combining a translation of $-L$, and a boost with velocity $v$, ``moves'' the neutron forward in time of an amount $\gamma vL$, causing a decay, for large $L$. Ultimately, this is also the physical meaning of equation \eqref{hqpqk}: ``time evolution'' (left-hand side) is the result of combining a space translation and a boost (right-hand side) \cite{Keister1991}.

Now, to recover equation \eqref{lambdasblurs}, we can just invoke the linearity of $U(\Lambda)$, and make the identification
\begin{equation}
\ket{\psi} \equiv \dfrac{\ket{\beta} + \ket{\alpha}}{\sqrt{2}} \, .
\end{equation}
Also, note that the event $(0,-L)$ occurs, in Bob's frame, at time $t_B=-\gamma v L$, so that \eqref{gvL} is geometrically  consistent with the Minkowski diagram in figure \ref{fig:boxes}.

\textit{Sometimes the decay is inescapable -} In the previous section, we ``cheated'' a bit. In fact, we considered a state $\ket{\alpha}$ that models a neutron ``in the origin''. But there is a problem: there is always a little uncertainty $\Delta x^1$ about the position of a particle. Therefore, when we boost \textit{any} wavefunction, its tails (no matter how short) are pushed one to the future and the other to the past, causing a little decay. How important is this effect?

\begin{figure}
\begin{center}
\includegraphics[width=0.49\textwidth]{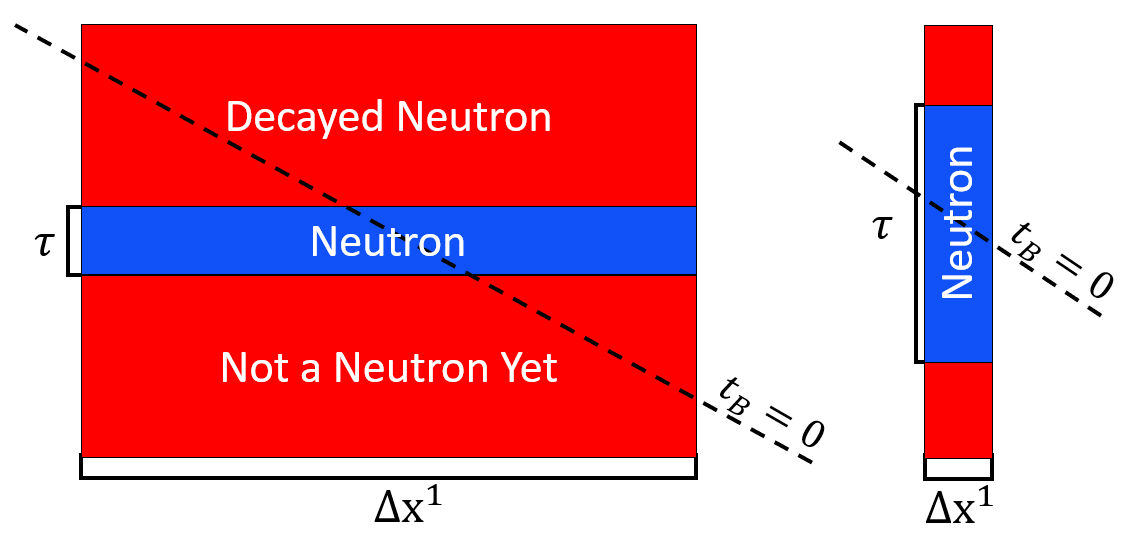}
	\caption{Minkowski diagrams of the wavefunction of a neutron for two different values of the ratio $\Delta x^1/\tau$. Left diagram: if $\Delta x^1 \gg \tau$, the hypersurface $\{t_B=0 \}$ of a moving observer (dashed line) covers a large region of spacetime where the particle is not a neutron (in red), so that $\bra{\Lambda\alpha}\mathcal{Q}\ket{\Lambda\alpha} \ll 1$. Right diagram: the condition $\Delta x^1 \ll \tau$ restores $\bra{\Lambda\alpha}\mathcal{Q}\ket{\Lambda\alpha} \approx 1$, because the hypersurface $\{t_B=0 \}$ intersects only the ``neutron region'' (in blue).}
	\label{fig:triple77}
	\end{center}
\end{figure}

Suppose that $\ket{\alpha}$ is a neutron wavepacket with center in the origin and zero average velocity. By contraction of lengths, the tails of the wavefunction $\ket{\Lambda\alpha}$ extend till $|x^1| \sim \Delta x^1/\gamma$. To estimate the tail ``desynchronization'', we can just evaluate equation \eqref{magnetic} on one tail (for $t=0$):
\begin{equation}\label{desincrocnizzo}
    \bra{\Lambda\alpha}j^{\mu}(0,\Delta x^1/\gamma)\ket{\Lambda\alpha} = \Lambda \indices{^\mu _\rho} \bra{\alpha}j^{\rho}( - v \Delta  x^1, \Delta x^1 )\ket{\alpha} \, .
\end{equation}
Comparing the times at which $j$ is evaluated, we can conclude that the desynchronization timescale between $\ket{\alpha}$ and $\ket{\Lambda\alpha}$ is $\Delta t \sim v \, \Delta x^1$. If this timescale becomes comparable to the decay time $\tau$, the neutron decays along the tails of the wavefunction, just by relativity of simultaneity. To avoid this possibility for all values of $v$, we must require that (see also figure \ref{fig:triple77})
\begin{equation}\label{detla}
 \Delta x^1 \ll \tau \, .
\end{equation}
This is the central inequality of the paper: when it is respected, one has $\bra{\Lambda\alpha}\mathcal{Q}\ket{\Lambda\alpha} \approx 1$, provided that the wavepacket is centered in the origin. This is also confirmed by the explicit calculation of \citet{Stefanovich:2005ai}. However, if \eqref{detla} is broken, a boost causes a measurable decay, no matter where we set the origin! Of course, an example of a state that violates \eqref{detla} is the state $\ket{\psi}$ of our thought experiment (see figure \ref{fig:boxes}).

\begin{figure*}
\begin{center}
\includegraphics[width=0.8\textwidth]{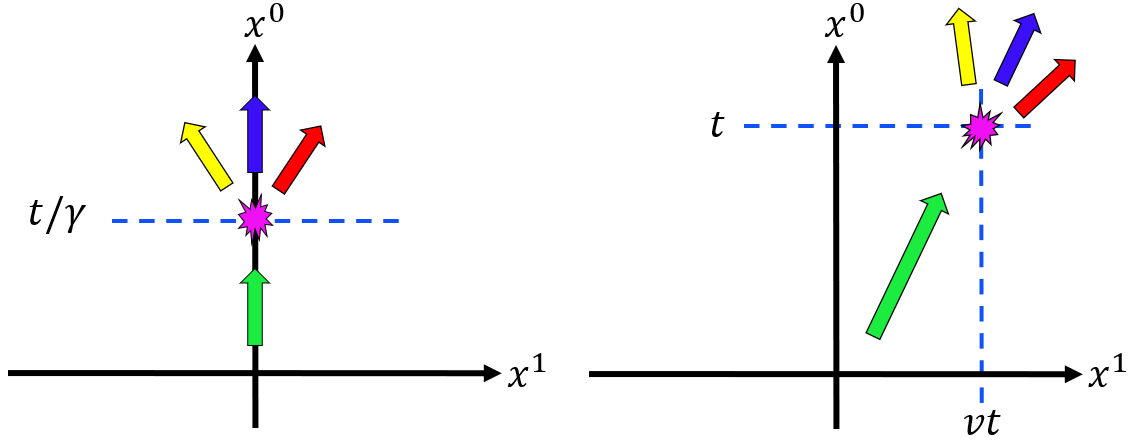}
	\caption{Minkowski diagrams of $ \bra{\beta} e^{-iHt/\gamma}  \ket{\alpha}$ (left panel) and $\bra{\Lambda \beta} e^{iP^1 vt} e^{-iHt} \ket{\Lambda \alpha}$ (right panel). The two processes are mapped into each other by a Lorentz boost. As a consequence, the value of the probability amplitude is exactly the same. Everything goes as Special Relativity predicts: when we boost, $t/\gamma$ is ``stretched'' into $t$ (time dilation). The translation $e^{i\hat{P}^1 vt}$ is necessary: the boosted neutron travels a distance $vt$ before decaying, so that we must project on $e^{-i\hat{P}^1 vt}\ket{\Lambda\beta}$, not on $\ket{\Lambda\beta}$. }
	\label{fig:triple}
	\end{center}
\end{figure*}

In the Appendix, we compute explicitly the average $\bra{\Lambda\alpha}\mathcal{Q}\ket{\Lambda\alpha}$, under the assumption that $\ket{\alpha}$ is a Gaussian wavepacket (at rest in the origin) with $\bra{\alpha}\mathcal{Q}\ket{\alpha}=1$. We obtain the approximate formula below:
\begin{equation}\label{suppleme}
\bra{\Lambda\alpha}\mathcal{Q}\ket{\Lambda\alpha} \approx e^{a^2} \text{erfc}|a| \quad \text{with }a:= \dfrac{v\Delta x^1}{\tau \sqrt{2}} \, ,
\end{equation}
where ``$\,\text{erfc}\,$'' is the complementary error function. As expected, if \eqref{detla} is obeyed (i.e. $a \rightarrow 0$), the above expression converges to $1$. But if \eqref{detla} is violated, then $\bra{\Lambda\alpha}\mathcal{Q}\ket{\Lambda\alpha} \sim a^{-1}$, which tends to zero for large $a$. 

Now, there is an important fact to note. A particle cannot be localised in a region of space that is smaller than the Compton wavelength $M^{-1}$ \cite{landau4,Kaloyerou1988,Eberhard1988,Barat2003}. Therefore, a single-particle state that obeys \eqref{detla} is allowed to exist \textit{only if} $M^{-1} \ll \tau$. As a consequence, we can always find two observers that disagree on whether a ``resonance particle'', with $M^{-1} \sim \tau$ (i.e. $M \sim  \Gamma := \tau^{-1}$ \cite{Alavi:2014mxa}), exists or not. In other words, the inequality $M^{-1} \ll \tau$ is a necessary (but not sufficient!) condition for establishing the (approximate) Lorentz-invariance of $\braket{\mathcal{Q}}$.

\textit{Quantum deviations from time dilation -} We are finally able to discuss the problem of time dilation. We first summarize the state of the art. Let $\ket{\chi}$ be an isolated neutron in an arbitrary state of motion. Since it is a neutron with probability $1$, we know that $\bra{\chi}\mathcal{Q}\ket{\chi}=1$. We let it evolve for a time $t$. The state now is $e^{-iHt}\ket{\chi}$, and the probability that we still have a neutron is 
\begin{equation}
\mathcal{P}(t) = \bra{\chi}e^{iHt}\mathcal{Q}e^{-iHt}\ket{\chi} \, .
\end{equation}
Since $\mathcal{Q}$ and $P^j$ are commuting observables, they can be diagonalised simultaneously. Thus, there is a set of neutron momentum eigenstates $\ket{n,\textbf{p},\sigma}$ such that ($\sigma$ is the spin)
\begin{equation}
\begin{split}
 {\mathcal{Q}} \ket{n,\textbf{p},\sigma} ={}&  \ket{n,\textbf{p},\sigma} \, , \\
 {P}^j \ket{n,\textbf{p},\sigma} ={}& p^j \ket{n,\textbf{p},\sigma} \, . \\ 
\end{split}
\end{equation}
One can expand $\mathcal{Q}$ and $\ket{\chi}$ using these states. All that remains is to calculate the characteristic amplitudes
\begin{equation}\label{ampliomoMMo}
\mathcal{A}(t)= \bra{n,\textbf{p},\sigma} e^{-iHt} \ket{n,\textbf{p},\sigma} \, .
\end{equation}
Let us jump directly to the result. Depending on the level of detail, the exact formula may change slightly, but all authors agree \cite{shirokov2005,stefanovich2006,Urbanowski:2014gza,Giacosa:2015mpm} that the decay timescale of a neutron with momentum $\textbf{p}$ can be expressed as
\begin{equation}\label{tauvcorreggo}
\tau(\textbf{p}) = \tau \, \dfrac{\sqrt{M^2 +\textbf{p}^2}}{M} \bigg[ 1+ \mathcal{O}\bigg(\dfrac{M^{-1}}{\tau}\bigg) \bigg] \, .
\end{equation}
Outside the bracket, we have the usual time-dilated decay time ``$\, \tau \gamma \,$''. The bracket is a pure quantum correction, which deviates from 1 only when the Compton wavelength $M^{-1}$ is comparable to the rest-frame decay time $\tau$. This correction has been a source of debate for a long time: is it just a mathematical artefact \cite{Alavi:2014mxa}, or are we observing a breakdown of Special Relativity (SR) \cite{Stefanovich:2005ai}? As we are going to show, neither. This effect is physical, and it does not contradict SR.

First, let us consider the identity below:
\begin{equation}
 e^{-iHt/\gamma} = U^\dagger(\Lambda) e^{iP^1 vt} e^{-iHt} U(\Lambda) \, .
\end{equation}
It can be easily proved by inverting equation \eqref{poincorro}. Its matrix element between two generic states $\bra{\beta}$ and $\ket{\alpha}$ is
\begin{equation}\label{sbuif}
 \bra{\beta} e^{-iHt/\gamma}  \ket{\alpha} = \bra{\Lambda \beta} e^{iP^1 vt} e^{-iHt} \ket{\Lambda \alpha} \, .
\end{equation}
To understand the physical meaning of this \textit{exact} identity, consider the case in which $\ket{\alpha}$ is a neutron at rest near the origin, and $\ket{\beta}$ is a triplet $p^+ + e^- + \bar{\nu}_e$. Then, the amplitudes above can be schematically plotted in a ``Feynman-Minkowski'' diagram, as in figure \ref{fig:triple}. As we can see, the phenomenon of time dilation is perfectly well captured by the quantum theory: the amplitude for $\ket{\alpha}$ to transform into $\ket{\beta} $ in a time $t/\gamma$ is equal to the amplitude for $\ket{\Lambda\alpha}$ to transform into $e^{-i\hat{P}^1 vt}\ket{\Lambda\beta} $ in a (longer) time $t$. There is no ``quantum breakdown of SR''.

However, there is a complication. If $M^{-1} \gtrsim \tau$, then it is impossible for $\ket{\alpha}$ to obey \eqref{detla} without breaking the Compton limit ($\Delta x^1 \gtrsim M^{-1}$). As a result, if $\ket{\alpha}$ is a neutron, in the sense that $\bra{\alpha}\mathcal{Q}\ket{\alpha}=1$, then in general $\ket{\Lambda\alpha}$ is \textit{not} a perfect neutron: $\bra{\Lambda\alpha}\mathcal{Q}\ket{\Lambda\alpha}<1$. Thus, we cannot construct moving neutron states by applying Lorentz boosts to neutrons at rest. Vice versa, if we take a moving neutron, and we boost to its rest frame, we no longer have a neutron. This implies that the mathematical identity \eqref{sbuif} cannot be used to relate the decay amplitude of a neutron at rest with that of a neutron in motion, because $\ket{\Lambda\alpha}$ (the ``boosted neutron'') and $\ket{\chi}$ (the moving neutron) are different states!

On the other hand, if $M^{-1} \ll \tau$, then it is possible to construct couples of states $\ket{\alpha}$ and $\ket{\Lambda\alpha}$ that are both neutrons, because \eqref{detla} does not violate the Compton limit. Now yes: we can use \eqref{sbuif} to relate the decay amplitudes for neutrons in different states of motion, and time dilation must be restored. This explains why the quantum correction in \eqref{tauvcorreggo} tends to zero in this limit. Indeed, in their derivation of time dilation, \citet{Alavi:2014mxa} were forced to assume that $M^{-1} \ll \Delta x^1 \ll \tau$. 

In conclusion, equation \eqref{suppleme} acts as a bridge between the analysis of \citet{Alavi:2014mxa} and the theorem of \citet{Exner:1983xu} and \citet{Stefanovich:2005ai}. In fact, in the regime considered by Alavi-Giunti (namely, $\Delta x^1 \ll \tau$), equation \eqref{suppleme} reduces to  $\bra{\Lambda\alpha}\mathcal{Q}\ket{\Lambda\alpha} \approx 1$, restoring our intuition that a neutron is perceived as a neutron in all reference frames. On the other hand, when $\Delta x^1$ becomes comparable to $\tau$, the operator $\mathcal{Q}$ ceases to be Lorentz-invariant, and $\bra{\Lambda\alpha}\mathcal{Q}\ket{\Lambda\alpha} <1$, in agreement with the Exner-Stefanovich theorem.

\textit{Application 1: neutrino decay -} Massive neutrinos may decay, and there are many possible decay channels outside the standard model \cite{Gomes2015}. As a proof of principle, let us see what happens if the heaviest neutrino has mass $M=0.05 \,$eV and rest-frame decay time $\tau = 10^{-13}\,$s (such extremely short lifetime is consistent with observational constraints  \cite{Gomes2015}). With the choice of parameters above, we get $M^{-1}/\tau \approx 0.83$. Hence, time dilation breaks down completely [see equation \eqref{tauvcorreggo}]. This is a serious problem, because all constraints on the neutrino lifetime assume from the start that time dilation is valid \cite{Baerwald2012}. Furthermore, if we set $\Delta x^1 \approx M^{-1}$ and $v \approx 1$ in \eqref{suppleme}, we obtain $\bra{\Lambda\alpha}\mathcal{Q}\ket{\Lambda\alpha} \approx 0.57$, meaning that a boosted neutrino has only $57 \%$ of probability of existing as a neutrino!
 
\textit{Application 2: sterile neutrinos -} \citet{Moss2018} consider a hypothetical sterile neutrino species, $\nu_4$, with very short lifetime: $\tau \approx 10^{-16} \,$s. Taking again $v \sim 1$ and $\Delta x^1 \approx M^{-1}$, and assuming $M = 1 \,$eV, we obtain $\bra{\Lambda\alpha}\mathcal{Q}\ket{\Lambda\alpha} \approx 0.02$. This means that such sterile neutrino disappears almost completely when we boost it. 

\textit{Application 3: boosting pions -} The lifetime of the neutral pion, $\pi^0$, is $\tau \approx 8.5 \times 10^{-17} \,$s \cite{ParticleDataGroup:2020ssz}. If we apply an ultra-relativistic boost ($v \sim 1$) to a wavefunction having rest-frame position uncertainty $\Delta x^1 \approx 5 \times 10^{-8} \,$m, we obtain $\bra{\Lambda\alpha}\mathcal{Q}\ket{\Lambda\alpha} \approx 0.34$. Whether values of $\Delta x^1$ of the order of $10^{-8} \,$m are actually attained in experiments will be subject of future investigation (note that in the laboratory frame the position uncertainty is shorter of a factor $1/\gamma$ \footnote{For bottonomium decays, the produced pions can have $\gamma \sim 50$. Hence, $\Delta x^1 \approx 5 \times 10^{-8} \,$m corresponds to $(\Delta x^1)_{\text{lab}} \approx  1 \,$nm in the laboratory frame.}). But if this happens, an ultra-relativistic $\pi^0$ exists only with probability $\sim 1/3$, provided that its existence probability is $1$ in the rest frame.

\textit{Future perspectives -} The role that relativity of simultaneity can play in the quantum dynamics of an unstable system has been overlooked till now \footnote{There is a similar problem also in relativistic hydrodynamics: many hydrodynamic theories end up being unstable, if the implications of relativity of simultaneity are not considered carefully \cite{Hiscock_Insatibility_first_order,GavassinoLyapunov_2020,GavassinoUEIT2021,GavassinoSuperluminal2021}.}. Here, we were focusing on what happens when we boost a single unstable particle. However, also larger systems should exhibit such counterintuitive effects. It would be interesting to apply this same set of ideas to an unstable field \cite{Lima2013}, and see if similar ``paradoxes'' occur. We leave this as a subject of future investigation.

\riga \\
L.G. acknowledges support by the Polish National Science Centre grant OPUS 2019/33/B/ST9/00942. 
F. G. acknowledge support from the Polish National Science Centre grant OPUS 2019/33/B/ST2/00613.

\bibliography{Biblio}

\begin{thebibliography}{66}%
\makeatletter
\providecommand \@ifxundefined [1]{%
 \@ifx{#1\undefined}
}%
\providecommand \@ifnum [1]{%
 \ifnum #1\expandafter \@firstoftwo
 \else \expandafter \@secondoftwo
 \fi
}%
\providecommand \@ifx [1]{%
 \ifx #1\expandafter \@firstoftwo
 \else \expandafter \@secondoftwo
 \fi
}%
\providecommand \natexlab [1]{#1}%
\providecommand \enquote  [1]{``#1''}%
\providecommand \bibnamefont  [1]{#1}%
\providecommand \bibfnamefont [1]{#1}%
\providecommand \citenamefont [1]{#1}%
\providecommand \href@noop [0]{\@secondoftwo}%
\providecommand \href [0]{\begingroup \@sanitize@url \@href}%
\providecommand \@href[1]{\@@startlink{#1}\@@href}%
\providecommand \@@href[1]{\endgroup#1\@@endlink}%
\providecommand \@sanitize@url [0]{\catcode `\\12\catcode `\$12\catcode
  `\&12\catcode `\#12\catcode `\^12\catcode `\_12\catcode `\%12\relax}%
\providecommand \@@startlink[1]{}%
\providecommand \@@endlink[0]{}%
\providecommand \url  [0]{\begingroup\@sanitize@url \@url }%
\providecommand \@url [1]{\endgroup\@href {#1}{\urlprefix }}%
\providecommand \urlprefix  [0]{URL }%
\providecommand \Eprint [0]{\href }%
\providecommand \doibase [0]{http://dx.doi.org/}%
\providecommand \selectlanguage [0]{\@gobble}%
\providecommand \bibinfo  [0]{\@secondoftwo}%
\providecommand \bibfield  [0]{\@secondoftwo}%
\providecommand \translation [1]{[#1]}%
\providecommand \BibitemOpen [0]{}%
\providecommand \bibitemStop [0]{}%
\providecommand \bibitemNoStop [0]{.\EOS\space}%
\providecommand \EOS [0]{\spacefactor3000\relax}%
\providecommand \BibitemShut  [1]{\csname bibitem#1\endcsname}%
\let\auto@bib@innerbib\@empty
\bibitem [{\citenamefont {{Zwanziger}}(1963)}]{Zwanziger1963}%
  \BibitemOpen
  \bibfield  {author} {\bibinfo {author} {\bibfnamefont {D.}~\bibnamefont
  {{Zwanziger}}},\ }\href {\doibase 10.1103/PhysRev.131.2818} {\bibfield
  {journal} {\bibinfo  {journal} {Physical Review}\ }\textbf {\bibinfo {volume}
  {131}},\ \bibinfo {pages} {2818} (\bibinfo {year} {1963})}\BibitemShut
  {NoStop}%
\bibitem [{\citenamefont {{Kawai}}\ and\ \citenamefont
  {{Got{\={o}}}}(1969)}]{Kawai1969}%
  \BibitemOpen
  \bibfield  {author} {\bibinfo {author} {\bibfnamefont {T.}~\bibnamefont
  {{Kawai}}}\ and\ \bibinfo {author} {\bibfnamefont {M.}~\bibnamefont
  {{Got{\={o}}}}},\ }\href {\doibase 10.1007/BF02712348} {\bibfield  {journal}
  {\bibinfo  {journal} {Nuovo Cimento B Serie}\ }\textbf {\bibinfo {volume}
  {60}},\ \bibinfo {pages} {21} (\bibinfo {year} {1969})}\BibitemShut {NoStop}%
\bibitem [{\citenamefont {{Weldon}}(1976)}]{Weldon1976}%
  \BibitemOpen
  \bibfield  {author} {\bibinfo {author} {\bibfnamefont {H.~A.}\ \bibnamefont
  {{Weldon}}},\ }\href {\doibase 10.1103/PhysRevD.14.2030} {\bibfield
  {journal} {\bibinfo  {journal} {\prd}\ }\textbf {\bibinfo {volume} {14}},\
  \bibinfo {pages} {2030} (\bibinfo {year} {1976})}\BibitemShut {NoStop}%
\bibitem [{\citenamefont {{Alicki}}\ \emph {et~al.}(1986)\citenamefont
  {{Alicki}}, \citenamefont {{Fannes}},\ and\ \citenamefont
  {{Verbeure}}}]{Alicki1986}%
  \BibitemOpen
  \bibfield  {author} {\bibinfo {author} {\bibfnamefont {R.}~\bibnamefont
  {{Alicki}}}, \bibinfo {author} {\bibfnamefont {M.}~\bibnamefont {{Fannes}}},
  \ and\ \bibinfo {author} {\bibfnamefont {A.}~\bibnamefont {{Verbeure}}},\
  }\href {\doibase 10.1088/0305-4470/19/6/021} {\bibfield  {journal} {\bibinfo
  {journal} {Journal of Physics A Mathematical General}\ }\textbf {\bibinfo
  {volume} {19}},\ \bibinfo {pages} {919} (\bibinfo {year} {1986})}\BibitemShut
  {NoStop}%
\bibitem [{\citenamefont {{Stefanovich}}(1996)}]{Stefanovich1996}%
  \BibitemOpen
  \bibfield  {author} {\bibinfo {author} {\bibfnamefont {E.~V.}\ \bibnamefont
  {{Stefanovich}}},\ }\href {\doibase 10.1007/BF02085762} {\bibfield  {journal}
  {\bibinfo  {journal} {International Journal of Theoretical Physics}\ }\textbf
  {\bibinfo {volume} {35}},\ \bibinfo {pages} {2539} (\bibinfo {year}
  {1996})}\BibitemShut {NoStop}%
\bibitem [{\citenamefont {{Shirokov}}(2004)}]{shirokov2004}%
  \BibitemOpen
  \bibfield  {author} {\bibinfo {author} {\bibfnamefont {M.}~\bibnamefont
  {{Shirokov}}},\ }\href {\doibase 10.1023/B:IJTP.0000048637.97460.87}
  {\bibfield  {journal} {\bibinfo  {journal} {International Journal of
  Theoretical Physics}\ }\textbf {\bibinfo {volume} {43}},\ \bibinfo {pages}
  {1541} (\bibinfo {year} {2004})}\BibitemShut {NoStop}%
\bibitem [{\citenamefont {{Shirokov}}(2005)}]{shirokov2005}%
  \BibitemOpen
  \bibfield  {author} {\bibinfo {author} {\bibfnamefont {M.~I.}\ \bibnamefont
  {{Shirokov}}},\ }\href@noop {} {\bibfield  {journal} {\bibinfo  {journal}
  {arXiv e-prints}\ ,\ \bibinfo {eid} {quant-ph/0508087}} (\bibinfo {year}
  {2005})},\ \Eprint {http://arxiv.org/abs/quant-ph/0508087}
  {arXiv:quant-ph/0508087 [quant-ph]} \BibitemShut {NoStop}%
\bibitem [{\citenamefont {{Stefanovich}}(2006)}]{stefanovich2006}%
  \BibitemOpen
  \bibfield  {author} {\bibinfo {author} {\bibfnamefont {E.~V.}\ \bibnamefont
  {{Stefanovich}}},\ }\href@noop {} {\bibfield  {journal} {\bibinfo  {journal}
  {arXiv e-prints}\ ,\ \bibinfo {eid} {physics/0603043}} (\bibinfo {year}
  {2006})},\ \Eprint {http://arxiv.org/abs/physics/0603043}
  {arXiv:physics/0603043 [physics.gen-ph]} \BibitemShut {NoStop}%
\bibitem [{\citenamefont {Alavi}\ and\ \citenamefont
  {Giunti}(2015)}]{Alavi:2014mxa}%
  \BibitemOpen
  \bibfield  {author} {\bibinfo {author} {\bibfnamefont {S.~A.}\ \bibnamefont
  {Alavi}}\ and\ \bibinfo {author} {\bibfnamefont {C.}~\bibnamefont {Giunti}},\
  }\href {\doibase 10.1209/0295-5075/109/60001} {\bibfield  {journal} {\bibinfo
   {journal} {EPL}\ }\textbf {\bibinfo {volume} {109}},\ \bibinfo {pages}
  {60001} (\bibinfo {year} {2015})},\ \Eprint {http://arxiv.org/abs/1412.3346}
  {arXiv:1412.3346 [quant-ph]} \BibitemShut {NoStop}%
\bibitem [{\citenamefont {Urbanowski}(2014)}]{Urbanowski:2014gza}%
  \BibitemOpen
  \bibfield  {author} {\bibinfo {author} {\bibfnamefont {K.}~\bibnamefont
  {Urbanowski}},\ }\href {\doibase 10.1016/j.physletb.2014.08.073} {\bibfield
  {journal} {\bibinfo  {journal} {Phys. Lett. B}\ }\textbf {\bibinfo {volume}
  {737}},\ \bibinfo {pages} {346} (\bibinfo {year} {2014})},\ \Eprint
  {http://arxiv.org/abs/1408.6564} {arXiv:1408.6564 [hep-ph]} \BibitemShut
  {NoStop}%
\bibitem [{\citenamefont {{Urbanowski}}\ and\ \citenamefont
  {{Raczy{\'n}ska}}(2014)}]{UrbaRazy2014}%
  \BibitemOpen
  \bibfield  {author} {\bibinfo {author} {\bibfnamefont {K.}~\bibnamefont
  {{Urbanowski}}}\ and\ \bibinfo {author} {\bibfnamefont {K.}~\bibnamefont
  {{Raczy{\'n}ska}}},\ }\href {\doibase 10.1016/j.physletb.2014.02.043}
  {\bibfield  {journal} {\bibinfo  {journal} {Physics Letters B}\ }\textbf
  {\bibinfo {volume} {731}},\ \bibinfo {pages} {236} (\bibinfo {year}
  {2014})},\ \Eprint {http://arxiv.org/abs/1303.6975} {arXiv:1303.6975
  [astro-ph.HE]} \BibitemShut {NoStop}%
\bibitem [{\citenamefont {Urbanowski}(2017)}]{Urbanowski:2017haf}%
  \BibitemOpen
  \bibfield  {author} {\bibinfo {author} {\bibfnamefont {K.}~\bibnamefont
  {Urbanowski}},\ }\href {\doibase 10.5506/APhysPolB.48.1847} {\bibfield
  {journal} {\bibinfo  {journal} {Acta Phys. Polon. B}\ }\textbf {\bibinfo
  {volume} {48}},\ \bibinfo {pages} {1847} (\bibinfo {year} {2017})},\ \Eprint
  {http://arxiv.org/abs/1711.06096} {arXiv:1711.06096 [physics.gen-ph]}
  \BibitemShut {NoStop}%
\bibitem [{\citenamefont {Exner}(1983)}]{Exner:1983xu}%
  \BibitemOpen
  \bibfield  {author} {\bibinfo {author} {\bibfnamefont {P.}~\bibnamefont
  {Exner}},\ }\href {\doibase 10.1103/PhysRevD.28.2621} {\bibfield  {journal}
  {\bibinfo  {journal} {Phys. Rev. D}\ }\textbf {\bibinfo {volume} {28}},\
  \bibinfo {pages} {2621} (\bibinfo {year} {1983})}\BibitemShut {NoStop}%
\bibitem [{\citenamefont {{Havl{\'\i}{\v{c}}ek}}\ and\ \citenamefont
  {{Exner}}(1973)}]{ExnerCheck1973}%
  \BibitemOpen
  \bibfield  {author} {\bibinfo {author} {\bibfnamefont {M.}~\bibnamefont
  {{Havl{\'\i}{\v{c}}ek}}}\ and\ \bibinfo {author} {\bibfnamefont
  {P.}~\bibnamefont {{Exner}}},\ }\href {\doibase 10.1007/BF01593909}
  {\bibfield  {journal} {\bibinfo  {journal} {Czechoslovak Journal of Physics}\
  }\textbf {\bibinfo {volume} {23}},\ \bibinfo {pages} {594} (\bibinfo {year}
  {1973})}\BibitemShut {NoStop}%
\bibitem [{\citenamefont {Stefanovich}(2005)}]{Stefanovich:2005ai}%
  \BibitemOpen
  \bibfield  {author} {\bibinfo {author} {\bibfnamefont {E.~V.}\ \bibnamefont
  {Stefanovich}},\ }\href@noop {} {\  (\bibinfo {year} {2005})},\ \Eprint
  {http://arxiv.org/abs/physics/0504062} {arXiv:physics/0504062} \BibitemShut
  {NoStop}%
\bibitem [{\citenamefont {Giacosa}(2016)}]{Giacosa:2015mpm}%
  \BibitemOpen
  \bibfield  {author} {\bibinfo {author} {\bibfnamefont {F.}~\bibnamefont
  {Giacosa}},\ }\href {\doibase 10.5506/APhysPolB.47.2135} {\bibfield
  {journal} {\bibinfo  {journal} {Acta Phys. Polon. B}\ }\textbf {\bibinfo
  {volume} {47}},\ \bibinfo {pages} {2135} (\bibinfo {year} {2016})},\ \Eprint
  {http://arxiv.org/abs/1512.00232} {arXiv:1512.00232 [hep-ph]} \BibitemShut
  {NoStop}%
\bibitem [{\citenamefont {Giacosa}(2018)}]{Giacosa:2018dzm}%
  \BibitemOpen
  \bibfield  {author} {\bibinfo {author} {\bibfnamefont {F.}~\bibnamefont
  {Giacosa}},\ }\href {\doibase 10.1155/2018/4672051} {\bibfield  {journal}
  {\bibinfo  {journal} {Adv. High Energy Phys.}\ }\textbf {\bibinfo {volume}
  {2018}},\ \bibinfo {pages} {4672051} (\bibinfo {year} {2018})},\ \Eprint
  {http://arxiv.org/abs/1804.02728} {arXiv:1804.02728 [hep-ph]} \BibitemShut
  {NoStop}%
\bibitem [{\citenamefont {Stefanovich}(2018)}]{Stefanovich:2018xji}%
  \BibitemOpen
  \bibfield  {author} {\bibinfo {author} {\bibfnamefont {E.~V.}\ \bibnamefont
  {Stefanovich}},\ }\href {\doibase 10.1155/2018/4657079} {\bibfield  {journal}
  {\bibinfo  {journal} {Adv. High Energy Phys.}\ }\textbf {\bibinfo {volume}
  {2018}},\ \bibinfo {pages} {4657079} (\bibinfo {year} {2018})},\ \Eprint
  {http://arxiv.org/abs/1801.01549} {arXiv:1801.01549 [physics.gen-ph]}
  \BibitemShut {NoStop}%
\bibitem [{\citenamefont {Bailey}\ \emph {et~al.}(1977)\citenamefont {Bailey}
  \emph {et~al.}}]{Bailey:1977de}%
  \BibitemOpen
  \bibfield  {author} {\bibinfo {author} {\bibfnamefont {J.}~\bibnamefont
  {Bailey}} \emph {et~al.},\ }\href {\doibase 10.1038/268301a0} {\bibfield
  {journal} {\bibinfo  {journal} {Nature}\ }\textbf {\bibinfo {volume} {268}},\
  \bibinfo {pages} {301} (\bibinfo {year} {1977})}\BibitemShut {NoStop}%
\bibitem [{\citenamefont {Bailey}\ \emph {et~al.}(1979)\citenamefont {Bailey}
  \emph {et~al.}}]{CERN-Mainz-Daresbury:1978ccd}%
  \BibitemOpen
  \bibfield  {author} {\bibinfo {author} {\bibfnamefont {J.}~\bibnamefont
  {Bailey}} \emph {et~al.} (\bibinfo {collaboration} {CERN-Mainz-Daresbury}),\
  }\href {\doibase 10.1016/0550-3213(79)90292-X} {\bibfield  {journal}
  {\bibinfo  {journal} {Nucl. Phys. B}\ }\textbf {\bibinfo {volume} {150}},\
  \bibinfo {pages} {1} (\bibinfo {year} {1979})}\BibitemShut {NoStop}%
\bibitem [{\citenamefont {{Baerwald}}\ \emph {et~al.}(2012)\citenamefont
  {{Baerwald}}, \citenamefont {{Bustamante}},\ and\ \citenamefont
  {{Winter}}}]{Baerwald2012}%
  \BibitemOpen
  \bibfield  {author} {\bibinfo {author} {\bibfnamefont {P.}~\bibnamefont
  {{Baerwald}}}, \bibinfo {author} {\bibfnamefont {M.}~\bibnamefont
  {{Bustamante}}}, \ and\ \bibinfo {author} {\bibfnamefont {W.}~\bibnamefont
  {{Winter}}},\ }\href {\doibase 10.1088/1475-7516/2012/10/020} {\bibfield
  {journal} {\bibinfo  {journal} {\jcap}\ }\textbf {\bibinfo {volume} {2012}},\
  \bibinfo {eid} {020} (\bibinfo {year} {2012})},\ \Eprint
  {http://arxiv.org/abs/1208.4600} {arXiv:1208.4600 [astro-ph.CO]} \BibitemShut
  {NoStop}%
\bibitem [{\citenamefont {{Lipari}}(2012)}]{Lipari2012}%
  \BibitemOpen
  \bibfield  {author} {\bibinfo {author} {\bibfnamefont {P.}~\bibnamefont
  {{Lipari}}},\ }\href {\doibase 10.1016/j.nima.2012.02.002} {\bibfield
  {journal} {\bibinfo  {journal} {Nuclear Instruments and Methods in Physics
  Research A}\ }\textbf {\bibinfo {volume} {692}},\ \bibinfo {pages} {106}
  (\bibinfo {year} {2012})}\BibitemShut {NoStop}%
\bibitem [{\citenamefont {{Farley}}(2015)}]{Farley2015}%
  \BibitemOpen
  \bibfield  {author} {\bibinfo {author} {\bibfnamefont {F.~J.~M.}\
  \bibnamefont {{Farley}}},\ }\href {\doibase 10.1142/9789814644150\_0015}
  {\bibfield  {journal} {\bibinfo  {journal} {60 YEARS OF CERN EXPERIMENTS AND
  DISCOVERIES. Edited by SCHOPPER HERWIG ET AL. Published by World Scientific
  Publishing Co. Pte. Ltd}\ }\textbf {\bibinfo {volume} {23}},\ \bibinfo
  {pages} {371} (\bibinfo {year} {2015})}\BibitemShut {NoStop}%
\bibitem [{\citenamefont {Schr\"{o}der}(2017)}]{SCHRODER2017}%
  \BibitemOpen
  \bibfield  {author} {\bibinfo {author} {\bibfnamefont {F.~G.}\ \bibnamefont
  {Schr\"{o}der}},\ }\href {\doibase
  https://doi.org/10.1016/j.ppnp.2016.12.002} {\bibfield  {journal} {\bibinfo
  {journal} {Progress in Particle and Nuclear Physics}\ }\textbf {\bibinfo
  {volume} {93}},\ \bibinfo {pages} {1} (\bibinfo {year} {2017})}\BibitemShut
  {NoStop}%
\bibitem [{\citenamefont {{Jaeckel}}\ \emph {et~al.}(2018)\citenamefont
  {{Jaeckel}}, \citenamefont {{Malta}},\ and\ \citenamefont
  {{Redondo}}}]{Jaeckel2018}%
  \BibitemOpen
  \bibfield  {author} {\bibinfo {author} {\bibfnamefont {J.}~\bibnamefont
  {{Jaeckel}}}, \bibinfo {author} {\bibfnamefont {P.~C.}\ \bibnamefont
  {{Malta}}}, \ and\ \bibinfo {author} {\bibfnamefont {J.}~\bibnamefont
  {{Redondo}}},\ }\href {\doibase 10.1103/PhysRevD.98.055032} {\bibfield
  {journal} {\bibinfo  {journal} {\prd}\ }\textbf {\bibinfo {volume} {98}},\
  \bibinfo {eid} {055032} (\bibinfo {year} {2018})},\ \Eprint
  {http://arxiv.org/abs/1702.02964} {arXiv:1702.02964 [hep-ph]} \BibitemShut
  {NoStop}%
\bibitem [{\citenamefont {{Joshipura}}\ \emph {et~al.}(2002)\citenamefont
  {{Joshipura}}, \citenamefont {{Mass{\'o}}},\ and\ \citenamefont
  {{Mohanty}}}]{Joshipura2002}%
  \BibitemOpen
  \bibfield  {author} {\bibinfo {author} {\bibfnamefont {A.~S.}\ \bibnamefont
  {{Joshipura}}}, \bibinfo {author} {\bibfnamefont {E.}~\bibnamefont
  {{Mass{\'o}}}}, \ and\ \bibinfo {author} {\bibfnamefont {S.}~\bibnamefont
  {{Mohanty}}},\ }\href {\doibase 10.1103/PhysRevD.66.113008} {\bibfield
  {journal} {\bibinfo  {journal} {\prd}\ }\textbf {\bibinfo {volume} {66}},\
  \bibinfo {eid} {113008} (\bibinfo {year} {2002})},\ \Eprint
  {http://arxiv.org/abs/hep-ph/0203181} {arXiv:hep-ph/0203181 [hep-ph]}
  \BibitemShut {NoStop}%
\bibitem [{\citenamefont {{Gomes}}\ \emph {et~al.}(2015)\citenamefont
  {{Gomes}}, \citenamefont {{Gomes}},\ and\ \citenamefont
  {{Peres}}}]{Gomes2015}%
  \BibitemOpen
  \bibfield  {author} {\bibinfo {author} {\bibfnamefont {R.~A.}\ \bibnamefont
  {{Gomes}}}, \bibinfo {author} {\bibfnamefont {A.~L.~G.}\ \bibnamefont
  {{Gomes}}}, \ and\ \bibinfo {author} {\bibfnamefont {O.~L.~G.}\ \bibnamefont
  {{Peres}}},\ }\href {\doibase 10.1016/j.physletb.2014.12.014} {\bibfield
  {journal} {\bibinfo  {journal} {Physics Letters B}\ }\textbf {\bibinfo
  {volume} {740}},\ \bibinfo {pages} {345} (\bibinfo {year} {2015})},\ \Eprint
  {http://arxiv.org/abs/1407.5640} {arXiv:1407.5640 [hep-ph]} \BibitemShut
  {NoStop}%
\bibitem [{\citenamefont {{Chacko}}\ \emph {et~al.}(2021)\citenamefont
  {{Chacko}}, \citenamefont {{Dev}}, \citenamefont {{Du}}, \citenamefont
  {{Poulin}},\ and\ \citenamefont {{Tsai}}}]{Chack2021}%
  \BibitemOpen
  \bibfield  {author} {\bibinfo {author} {\bibfnamefont {Z.}~\bibnamefont
  {{Chacko}}}, \bibinfo {author} {\bibfnamefont {A.}~\bibnamefont {{Dev}}},
  \bibinfo {author} {\bibfnamefont {P.}~\bibnamefont {{Du}}}, \bibinfo {author}
  {\bibfnamefont {V.}~\bibnamefont {{Poulin}}}, \ and\ \bibinfo {author}
  {\bibfnamefont {Y.}~\bibnamefont {{Tsai}}},\ }\href {\doibase
  10.1103/PhysRevD.103.043519} {\bibfield  {journal} {\bibinfo  {journal}
  {\prd}\ }\textbf {\bibinfo {volume} {103}},\ \bibinfo {eid} {043519}
  (\bibinfo {year} {2021})},\ \Eprint {http://arxiv.org/abs/2002.08401}
  {arXiv:2002.08401 [astro-ph.CO]} \BibitemShut {NoStop}%
\bibitem [{\citenamefont {Moss}\ \emph {et~al.}(2018)\citenamefont {Moss},
  \citenamefont {Moulai}, \citenamefont {Arg\"uelles},\ and\ \citenamefont
  {Conrad}}]{Moss2018}%
  \BibitemOpen
  \bibfield  {author} {\bibinfo {author} {\bibfnamefont {Z.}~\bibnamefont
  {Moss}}, \bibinfo {author} {\bibfnamefont {M.~H.}\ \bibnamefont {Moulai}},
  \bibinfo {author} {\bibfnamefont {C.~A.}\ \bibnamefont {Arg\"uelles}}, \ and\
  \bibinfo {author} {\bibfnamefont {J.~M.}\ \bibnamefont {Conrad}},\ }\href
  {\doibase 10.1103/PhysRevD.97.055017} {\bibfield  {journal} {\bibinfo
  {journal} {Phys. Rev. D}\ }\textbf {\bibinfo {volume} {97}},\ \bibinfo
  {pages} {055017} (\bibinfo {year} {2018})}\BibitemShut {NoStop}%
\bibitem [{\citenamefont {Wald}(1984)}]{Wald}%
  \BibitemOpen
  \bibfield  {author} {\bibinfo {author} {\bibfnamefont {R.~M.}\ \bibnamefont
  {Wald}},\ }\href {https://cds.cern.ch/record/106274} {\emph {\bibinfo {title}
  {{General relativity}}}}\ (\bibinfo  {publisher} {Chicago Univ. Press},\
  \bibinfo {address} {Chicago, IL},\ \bibinfo {year} {1984})\BibitemShut
  {NoStop}%
\bibitem [{\citenamefont {Gourgoulhon}(2013)}]{special_in_gen}%
  \BibitemOpen
  \bibfield  {author} {\bibinfo {author} {\bibfnamefont {E.}~\bibnamefont
  {Gourgoulhon}},\ }\href@noop {} {\emph {\bibinfo {title} {Special Relativity
  in General Frames: From Particles to Astrophysics}}},\ \bibinfo {edition}
  {1st}\ ed.,\ Graduate Texts in Physics\ (\bibinfo  {publisher}
  {Springer-Verlag Berlin Heidelberg},\ \bibinfo {year} {2013})\BibitemShut
  {NoStop}%
\bibitem [{\citenamefont {Gavassino}(2022)}]{GavassinoSuperluminal2021}%
  \BibitemOpen
  \bibfield  {author} {\bibinfo {author} {\bibfnamefont {L.}~\bibnamefont
  {Gavassino}},\ }\href {\doibase 10.1103/PhysRevX.12.041001} {\bibfield
  {journal} {\bibinfo  {journal} {Phys. Rev. X}\ }\textbf {\bibinfo {volume}
  {12}},\ \bibinfo {pages} {041001} (\bibinfo {year} {2022})}\BibitemShut
  {NoStop}%
\bibitem [{Note1()}]{Note1}%
  \BibitemOpen
  \bibinfo {note} {Note that the lifetime of the neutron is subject to
  uncertainties due to incompatible results obtained with experimental
  different methods \cite {Wietfeldt:2011suo}. It has been speculated that this
  anomaly is due to Beyon-Standard-Model physics \cite
  {Fornal:2018eol,Berezhiani:2018eds} or the Anti-Zeno effect \cite
  {Giacosa:2019nbz}.}\BibitemShut {Stop}%
\bibitem [{Note2()}]{Note2}%
  \BibitemOpen
  \bibinfo {note} {Equation \protect \textup {\hbox {\mathsurround \z@ \protect
  \normalfont (\ignorespaces \ref {magnetic}\unskip \@@italiccorr )}} is just
  the transformation law of a vector field \cite {MTW_book,Weinberg_book_1972}
  in Quantum Field Theory \cite {wightman_book,BjorkenDrell_book,
  nakanishi_book, Ticciati1999,Peskin_book}. For a rigorous proof of \protect
  \textup {\hbox {\mathsurround \z@ \protect \normalfont (\ignorespaces \ref
  {magnetic}\unskip \@@italiccorr )}} in the context of (fully interacting)
  quantum electrodynamics see appendix B of \protect \citet {Zumino1960}. Note
  that equation \protect \textup {\hbox {\mathsurround \z@ \protect \normalfont
  (\ignorespaces \ref {magnetic}\unskip \@@italiccorr )}} is valid also in the
  context of (fully interacting) Relativistic Quantum Dynamics, see equation
  (9.4) of \protect \citet {Keister1991}.}\BibitemShut {Stop}%
\bibitem [{\citenamefont {Jackson}(1977)}]{Jackson1977}%
  \BibitemOpen
  \bibfield  {author} {\bibinfo {author} {\bibfnamefont {J.~D.}\ \bibnamefont
  {Jackson}},\ }\href@noop {} {\bibfield  {journal} {\bibinfo  {journal} {CERN
  report}\ }\textbf {\bibinfo {volume} {77-17}} (\bibinfo {year}
  {1977})}\BibitemShut {NoStop}%
\bibitem [{\citenamefont {{Boyer}}(1988)}]{Boyer1988}%
  \BibitemOpen
  \bibfield  {author} {\bibinfo {author} {\bibfnamefont {T.~H.}\ \bibnamefont
  {{Boyer}}},\ }\href {\doibase 10.1119/1.15501} {\bibfield  {journal}
  {\bibinfo  {journal} {American Journal of Physics}\ }\textbf {\bibinfo
  {volume} {56}},\ \bibinfo {pages} {688} (\bibinfo {year} {1988})}\BibitemShut
  {NoStop}%
\bibitem [{\citenamefont {Mezei}(1988)}]{Mezei1988}%
  \BibitemOpen
  \bibfield  {author} {\bibinfo {author} {\bibfnamefont {F.}~\bibnamefont
  {Mezei}},\ }\href@noop {} {\bibfield  {journal} {\bibinfo  {journal} {Acta
  Physica Hungarica}\ }\textbf {\bibinfo {volume} {64}} (\bibinfo {year}
  {1988})}\BibitemShut {NoStop}%
\bibitem [{\citenamefont {Zyla}\ \emph {et~al.}(2020)\citenamefont {Zyla} \emph
  {et~al.}}]{ParticleDataGroup:2020ssz}%
  \BibitemOpen
  \bibfield  {author} {\bibinfo {author} {\bibfnamefont {P.~A.}\ \bibnamefont
  {Zyla}} \emph {et~al.} (\bibinfo {collaboration} {Particle Data Group}),\
  }\href {\doibase 10.1093/ptep/ptaa104} {\bibfield  {journal} {\bibinfo
  {journal} {PTEP}\ }\textbf {\bibinfo {volume} {2020}},\ \bibinfo {pages}
  {083C01} (\bibinfo {year} {2020})}\BibitemShut {NoStop}%
\bibitem [{Note3()}]{Note3}%
  \BibitemOpen
  \bibinfo {note} {All observers agree on the spacetime topology \cite
  {Hawking1973} of the support of a tensor field.}\BibitemShut {Stop}%
\bibitem [{\citenamefont {Weinberg}(1995)}]{weinbergQFT_1995}%
  \BibitemOpen
  \bibfield  {author} {\bibinfo {author} {\bibfnamefont {S.}~\bibnamefont
  {Weinberg}},\ }\href {\doibase 10.1017/CBO9781139644167} {\emph {\bibinfo
  {title} {The Quantum Theory of Fields}}},\ Vol.~\bibinfo {volume} {1}\
  (\bibinfo  {publisher} {Cambridge University Press},\ \bibinfo {year}
  {1995})\BibitemShut {NoStop}%
\bibitem [{\citenamefont {Srednicki}(2007)}]{srednicki_book}%
  \BibitemOpen
  \bibfield  {author} {\bibinfo {author} {\bibfnamefont {M.}~\bibnamefont
  {Srednicki}},\ }\href {\doibase 10.1017/CBO9780511813917} {\emph {\bibinfo
  {title} {Quantum Field Theory}}}\ (\bibinfo  {publisher} {Cambridge
  University Press},\ \bibinfo {year} {2007})\BibitemShut {NoStop}%
\bibitem [{\citenamefont {Keister}\ and\ \citenamefont
  {Polyzou}(1991)}]{Keister1991}%
  \BibitemOpen
  \bibfield  {author} {\bibinfo {author} {\bibfnamefont {B.~D.}\ \bibnamefont
  {Keister}}\ and\ \bibinfo {author} {\bibfnamefont {W.~N.}\ \bibnamefont
  {Polyzou}},\ }\href@noop {} {\bibfield  {journal} {\bibinfo  {journal} {Adv.
  Nucl. Phys.}\ }\textbf {\bibinfo {volume} {20}},\ \bibinfo {pages} {225}
  (\bibinfo {year} {1991})}\BibitemShut {NoStop}%
\bibitem [{\citenamefont {Berestetskii}\ \emph {et~al.}(1973)\citenamefont
  {Berestetskii}, \citenamefont {Lifshitz},\ and\ \citenamefont
  {Pitaevskii}}]{landau4}%
  \BibitemOpen
  \bibfield  {author} {\bibinfo {author} {\bibfnamefont {V.}~\bibnamefont
  {Berestetskii}}, \bibinfo {author} {\bibfnamefont {E.}~\bibnamefont
  {Lifshitz}}, \ and\ \bibinfo {author} {\bibfnamefont {L.}~\bibnamefont
  {Pitaevskii}},\ }\href@noop {} {\emph {\bibinfo {title} {Relativistic Quantum
  Theory}}},\ \bibinfo {number} {v. 4}\ (\bibinfo  {publisher} {Pergamon
  Press},\ \bibinfo {year} {1973})\BibitemShut {NoStop}%
\bibitem [{\citenamefont {{Kaloyerou}}(1988)}]{Kaloyerou1988}%
  \BibitemOpen
  \bibfield  {author} {\bibinfo {author} {\bibfnamefont {P.~N.}\ \bibnamefont
  {{Kaloyerou}}},\ }\href {\doibase 10.1016/0375-9601(88)90333-7} {\bibfield
  {journal} {\bibinfo  {journal} {Physics Letters A}\ }\textbf {\bibinfo
  {volume} {129}},\ \bibinfo {pages} {285} (\bibinfo {year}
  {1988})}\BibitemShut {NoStop}%
\bibitem [{\citenamefont {Eberhard}\ and\ \citenamefont
  {Ross}(1989)}]{Eberhard1988}%
  \BibitemOpen
  \bibfield  {author} {\bibinfo {author} {\bibfnamefont {P.~H.}\ \bibnamefont
  {Eberhard}}\ and\ \bibinfo {author} {\bibfnamefont {R.~R.}\ \bibnamefont
  {Ross}},\ }\href {\doibase 10.1007/BF00696109} {\bibfield  {journal}
  {\bibinfo  {journal} {Found. Phys.}\ }\textbf {\bibinfo {volume} {2}},\
  \bibinfo {pages} {127} (\bibinfo {year} {1989})}\BibitemShut {NoStop}%
\bibitem [{\citenamefont {{Barat}}\ and\ \citenamefont
  {{Kimball}}(2003)}]{Barat2003}%
  \BibitemOpen
  \bibfield  {author} {\bibinfo {author} {\bibfnamefont {N.}~\bibnamefont
  {{Barat}}}\ and\ \bibinfo {author} {\bibfnamefont {J.~C.}\ \bibnamefont
  {{Kimball}}},\ }\href {\doibase 10.1016/S0375-9601(02)01806-6} {\bibfield
  {journal} {\bibinfo  {journal} {Physics Letters A}\ }\textbf {\bibinfo
  {volume} {308}},\ \bibinfo {pages} {110} (\bibinfo {year} {2003})},\ \Eprint
  {http://arxiv.org/abs/quant-ph/0111060} {arXiv:quant-ph/0111060 [quant-ph]}
  \BibitemShut {NoStop}%
\bibitem [{Note4()}]{Note4}%
  \BibitemOpen
  \bibinfo {note} {For bottonomium decays, the produced pions can have $\gamma
  \sim 50$. Hence, $\Delta x^1 \approx 5 \times 10^{-8} \protect \tmspace
  +\thinmuskip {.1667em}$m corresponds to $(\Delta x^1)_{\protect \text {lab}}
  \approx 1 \protect \tmspace +\thinmuskip {.1667em}$nm in the laboratory
  frame.}\BibitemShut {Stop}%
\bibitem [{Note5()}]{Note5}%
  \BibitemOpen
  \bibinfo {note} {There is a similar problem also in relativistic
  hydrodynamics: many hydrodynamic theories end up being unstable, if the
  implications of relativity of simultaneity are not considered carefully \cite
  {Hiscock_Insatibility_first_order,GavassinoLyapunov_2020,GavassinoUEIT2021,GavassinoSuperluminal2021}.}\BibitemShut
  {Stop}%
\bibitem [{\citenamefont {Lima}(2013)}]{Lima2013}%
  \BibitemOpen
  \bibfield  {author} {\bibinfo {author} {\bibfnamefont {W.~C.~C.}\
  \bibnamefont {Lima}},\ }\href {\doibase 10.1103/PhysRevD.88.124005}
  {\bibfield  {journal} {\bibinfo  {journal} {Phys. Rev. D}\ }\textbf {\bibinfo
  {volume} {88}},\ \bibinfo {pages} {124005} (\bibinfo {year}
  {2013})}\BibitemShut {NoStop}%
\bibitem [{\citenamefont {Wietfeldt}\ and\ \citenamefont
  {Greene}(2011)}]{Wietfeldt:2011suo}%
  \BibitemOpen
  \bibfield  {author} {\bibinfo {author} {\bibfnamefont {F.~E.}\ \bibnamefont
  {Wietfeldt}}\ and\ \bibinfo {author} {\bibfnamefont {G.~L.}\ \bibnamefont
  {Greene}},\ }\href {\doibase 10.1103/RevModPhys.83.1173} {\bibfield
  {journal} {\bibinfo  {journal} {Rev. Mod. Phys.}\ }\textbf {\bibinfo {volume}
  {83}},\ \bibinfo {pages} {1173} (\bibinfo {year} {2011})}\BibitemShut
  {NoStop}%
\bibitem [{\citenamefont {Fornal}\ and\ \citenamefont
  {Grinstein}(2018)}]{Fornal:2018eol}%
  \BibitemOpen
  \bibfield  {author} {\bibinfo {author} {\bibfnamefont {B.}~\bibnamefont
  {Fornal}}\ and\ \bibinfo {author} {\bibfnamefont {B.}~\bibnamefont
  {Grinstein}},\ }\href {\doibase 10.1103/PhysRevLett.120.191801} {\bibfield
  {journal} {\bibinfo  {journal} {Phys. Rev. Lett.}\ }\textbf {\bibinfo
  {volume} {120}},\ \bibinfo {pages} {191801} (\bibinfo {year} {2018})},\
  \bibinfo {note} {[Erratum: Phys.Rev.Lett. 124, 219901 (2020)]},\ \Eprint
  {http://arxiv.org/abs/1801.01124} {arXiv:1801.01124 [hep-ph]} \BibitemShut
  {NoStop}%
\bibitem [{\citenamefont {Berezhiani}(2019)}]{Berezhiani:2018eds}%
  \BibitemOpen
  \bibfield  {author} {\bibinfo {author} {\bibfnamefont {Z.}~\bibnamefont
  {Berezhiani}},\ }\href {\doibase 10.1140/epjc/s10052-019-6995-x} {\bibfield
  {journal} {\bibinfo  {journal} {Eur. Phys. J. C}\ }\textbf {\bibinfo {volume}
  {79}},\ \bibinfo {pages} {484} (\bibinfo {year} {2019})},\ \Eprint
  {http://arxiv.org/abs/1807.07906} {arXiv:1807.07906 [hep-ph]} \BibitemShut
  {NoStop}%
\bibitem [{\citenamefont {Giacosa}\ and\ \citenamefont
  {Pagliara}(2020)}]{Giacosa:2019nbz}%
  \BibitemOpen
  \bibfield  {author} {\bibinfo {author} {\bibfnamefont {F.}~\bibnamefont
  {Giacosa}}\ and\ \bibinfo {author} {\bibfnamefont {G.}~\bibnamefont
  {Pagliara}},\ }\href {\doibase 10.1103/PhysRevD.101.056003} {\bibfield
  {journal} {\bibinfo  {journal} {Phys. Rev. D}\ }\textbf {\bibinfo {volume}
  {101}},\ \bibinfo {pages} {056003} (\bibinfo {year} {2020})},\ \Eprint
  {http://arxiv.org/abs/1906.10024} {arXiv:1906.10024 [hep-ph]} \BibitemShut
  {NoStop}%
\bibitem [{\citenamefont {{Misner}}\ \emph {et~al.}(1973)\citenamefont
  {{Misner}}, \citenamefont {{Thorne}},\ and\ \citenamefont
  {{Wheeler}}}]{MTW_book}%
  \BibitemOpen
  \bibfield  {author} {\bibinfo {author} {\bibfnamefont {C.~W.}\ \bibnamefont
  {{Misner}}}, \bibinfo {author} {\bibfnamefont {K.~S.}\ \bibnamefont
  {{Thorne}}}, \ and\ \bibinfo {author} {\bibfnamefont {J.~A.}\ \bibnamefont
  {{Wheeler}}},\ }\href@noop {} {\emph {\bibinfo {title} {San Francisco:
  W.H.~Freeman and Co., 1973}}}\ (\bibinfo {year} {1973})\BibitemShut {NoStop}%
\bibitem [{\citenamefont {{Weinberg}}(1972)}]{Weinberg_book_1972}%
  \BibitemOpen
  \bibfield  {author} {\bibinfo {author} {\bibfnamefont {S.}~\bibnamefont
  {{Weinberg}}},\ }\href@noop {} {\emph {\bibinfo {title} {{Gravitation and
  Cosmology: Principles and Applications of the General Theory of
  Relativity}}}}\ (\bibinfo {year} {1972})\BibitemShut {NoStop}%
\bibitem [{\citenamefont {Streater}\ and\ \citenamefont
  {Wightman}(1964)}]{wightman_book}%
  \BibitemOpen
  \bibfield  {author} {\bibinfo {author} {\bibfnamefont {R.~F.}\ \bibnamefont
  {Streater}}\ and\ \bibinfo {author} {\bibfnamefont {A.~S.}\ \bibnamefont
  {Wightman}},\ }\href@noop {} {\emph {\bibinfo {title} {PCT, spin and
  statistics, and all that}}}\ (\bibinfo  {publisher} {Princeton University
  Press},\ \bibinfo {year} {1964})\BibitemShut {NoStop}%
\bibitem [{\citenamefont {Bjorken}\ and\ \citenamefont
  {Drell}(1965)}]{BjorkenDrell_book}%
  \BibitemOpen
  \bibfield  {author} {\bibinfo {author} {\bibfnamefont {J.~D.}\ \bibnamefont
  {Bjorken}}\ and\ \bibinfo {author} {\bibfnamefont {S.~D.}\ \bibnamefont
  {Drell}},\ }\href@noop {} {\emph {\bibinfo {title} {Relativistic Quantum
  Fields}}}\ (\bibinfo  {publisher} {McGraw-Hill Book Company},\ \bibinfo
  {year} {1965})\BibitemShut {NoStop}%
\bibitem [{\citenamefont {{Nakanishi}}\ and\ \citenamefont
  {{Ojima}}(1990)}]{nakanishi_book}%
  \BibitemOpen
  \bibfield  {author} {\bibinfo {author} {\bibfnamefont {N.}~\bibnamefont
  {{Nakanishi}}}\ and\ \bibinfo {author} {\bibfnamefont {I.}~\bibnamefont
  {{Ojima}}},\ }\href@noop {} {\emph {\bibinfo {title} {{Covariant Operator
  Formalism of Gauge Theories and Quantum Gravity}}}}\ (\bibinfo  {publisher}
  {World Scientific},\ \bibinfo {year} {1990})\BibitemShut {NoStop}%
\bibitem [{\citenamefont {Ticciati}(1999)}]{Ticciati1999}%
  \BibitemOpen
  \bibfield  {author} {\bibinfo {author} {\bibfnamefont {R.}~\bibnamefont
  {Ticciati}},\ }\href@noop {} {\emph {\bibinfo {title} {{Quantum Field Theory
  for Mathematicians}}}},\ Encyclopedia of Mathematics and its Applications\
  (\bibinfo  {publisher} {Cambridge Univ. Press},\ \bibinfo {address}
  {Cambridge},\ \bibinfo {year} {1999})\BibitemShut {NoStop}%
\bibitem [{\citenamefont {Peskin}\ and\ \citenamefont
  {Schroeder}(1995)}]{Peskin_book}%
  \BibitemOpen
  \bibfield  {author} {\bibinfo {author} {\bibfnamefont {M.~E.}\ \bibnamefont
  {Peskin}}\ and\ \bibinfo {author} {\bibfnamefont {D.~V.}\ \bibnamefont
  {Schroeder}},\ }\href {http://www.slac.stanford.edu/~mpeskin/QFT.html} {\emph
  {\bibinfo {title} {{An introduction to quantum field theory}}}}\ (\bibinfo
  {publisher} {Addison-Wesley},\ \bibinfo {address} {Reading, USA},\ \bibinfo
  {year} {1995})\BibitemShut {NoStop}%
\bibitem [{\citenamefont {{Zumino}}(1960)}]{Zumino1960}%
  \BibitemOpen
  \bibfield  {author} {\bibinfo {author} {\bibfnamefont {B.}~\bibnamefont
  {{Zumino}}},\ }\href {\doibase 10.1063/1.1703632} {\bibfield  {journal}
  {\bibinfo  {journal} {Journal of Mathematical Physics}\ }\textbf {\bibinfo
  {volume} {1}},\ \bibinfo {pages} {1} (\bibinfo {year} {1960})}\BibitemShut
  {NoStop}%
\bibitem [{\citenamefont {Hawking}\ and\ \citenamefont
  {Ellis}(2011)}]{Hawking1973}%
  \BibitemOpen
  \bibfield  {author} {\bibinfo {author} {\bibfnamefont {S.~W.}\ \bibnamefont
  {Hawking}}\ and\ \bibinfo {author} {\bibfnamefont {G.~F.~R.}\ \bibnamefont
  {Ellis}},\ }\href {\doibase 10.1017/CBO9780511524646} {\emph {\bibinfo
  {title} {{The Large Scale Structure of Space-Time}}}},\ Cambridge Monographs
  on Mathematical Physics\ (\bibinfo  {publisher} {Cambridge University
  Press},\ \bibinfo {year} {2011})\BibitemShut {NoStop}%
\bibitem [{\citenamefont {Hiscock}\ and\ \citenamefont
  {Lindblom}(1985)}]{Hiscock_Insatibility_first_order}%
  \BibitemOpen
  \bibfield  {author} {\bibinfo {author} {\bibfnamefont {W.}~\bibnamefont
  {Hiscock}}\ and\ \bibinfo {author} {\bibfnamefont {L.}~\bibnamefont
  {Lindblom}},\ }\href {\doibase 10.1103/PhysRevD.31.725} {\bibfield  {journal}
  {\bibinfo  {journal} {Physical review D: Particles and fields}\ }\textbf
  {\bibinfo {volume} {31}},\ \bibinfo {pages} {725} (\bibinfo {year}
  {1985})}\BibitemShut {NoStop}%
\bibitem [{\citenamefont {Gavassino}\ \emph {et~al.}(2020)\citenamefont
  {Gavassino}, \citenamefont {Antonelli},\ and\ \citenamefont
  {Haskell}}]{GavassinoLyapunov_2020}%
  \BibitemOpen
  \bibfield  {author} {\bibinfo {author} {\bibfnamefont {L.}~\bibnamefont
  {Gavassino}}, \bibinfo {author} {\bibfnamefont {M.}~\bibnamefont
  {Antonelli}}, \ and\ \bibinfo {author} {\bibfnamefont {B.}~\bibnamefont
  {Haskell}},\ }\href {\doibase 10.1103/physrevd.102.043018} {\bibfield
  {journal} {\bibinfo  {journal} {Physical Review D}\ }\textbf {\bibinfo
  {volume} {102}} (\bibinfo {year} {2020}),\
  10.1103/physrevd.102.043018}\BibitemShut {NoStop}%
\bibitem [{\citenamefont {{Gavassino}}\ and\ \citenamefont
  {{Antonelli}}(2021)}]{GavassinoUEIT2021}%
  \BibitemOpen
  \bibfield  {author} {\bibinfo {author} {\bibfnamefont {L.}~\bibnamefont
  {{Gavassino}}}\ and\ \bibinfo {author} {\bibfnamefont {M.}~\bibnamefont
  {{Antonelli}}},\ }\href {\doibase 10.3389/fspas.2021.686344} {\bibfield
  {journal} {\bibinfo  {journal} {Frontiers in Astronomy and Space Sciences}\
  }\textbf {\bibinfo {volume} {8}},\ \bibinfo {eid} {92} (\bibinfo {year}
  {2021})},\ \Eprint {http://arxiv.org/abs/2105.15184} {arXiv:2105.15184
  [gr-qc]} \BibitemShut {NoStop}%
\bibitem [{Note6()}]{Note6}%
  \BibitemOpen
  \bibinfo {note} {The invariance of $\protect \mathcal {Q}$ under space
  translations is a direct consequence of equation \protect \textup {\hbox
  {\mathsurround \z@ \protect \normalfont (\ignorespaces \ref {QQQQQQQ}\unskip
  \@@italiccorr )}}. To see this, one can just invoke the well-known \cite
  {srednicki_book} identity $e^{iP^1a}J^0(t,x)e^{-iP^1 a}=J^0(t,x-a)$, and
  change the integration variable in \protect \textup {\hbox {\mathsurround \z@
  \protect \normalfont (\ignorespaces \ref {QQQQQQQ}\unskip \@@italiccorr )}}
  from $x$ to $x-a$, obtaining $e^{iP^1a}\protect \mathcal {Q} e^{-iP^1
  a}=\protect \mathcal {Q}$.}\BibitemShut {Stop}%
\end{thebibliography}%

\onecolumngrid

\newpage

\begin{center}
\textbf{\large APPENDIX}\\
\end{center}

\section{A simple formula}\label{simpluzza}


In this section, we derive a simple analytical formula for the probability $\mathcal{P}(t=0)=\bra{\Lambda\alpha} \mathcal{Q}\ket{\Lambda\alpha}$ that a boosted neutron (located nearby the origin) is still a neutron (at $t=0$). It is only a rough estimate,  but it is important to have an expression that can be used in ``back-of-the-envelope calculations''. For simplicity, we work in 1+1 dimensions. 

We start from a simple observation: if there is only one baryon, the projector $\mathcal{Q}$ may be interpreted as the ``neutron number'', namely an effective (non-conserved) charge that counts ``how many neutrons are there'', across all space. A quantum number of this kind can be expressed (see Section 15.8 of \citet{BjorkenDrell_book})  as the flux of some associated current $J^\nu$ through hyperplanes $\{\text{time} = \text{const} \}$, namely \footnote{The invariance of $\mathcal{Q}$ under space translations is a direct consequence of equation \eqref{QQQQQQQ}. To see this, one can just invoke the well-known \cite{srednicki_book} identity $e^{iP^1a}J^0(t,x)e^{-iP^1 a}=J^0(t,x-a)$, and change the integration variable in \eqref{QQQQQQQ} from $x$ to $x-a$, obtaining $e^{iP^1a}\mathcal{Q} e^{-iP^1 a}=\mathcal{Q}$.} 
\begin{equation}\label{QQQQQQQ}
\mathcal{Q} = \int_{t = \text{const}} \!\!\!\!\!\!\!\!\!  d\Sigma_\nu \, J^\nu = \int dx \, J^0 \, .
\end{equation}
The decay of the neutron is possible because $J^\nu$ is \textit{not} a Noether current of the field theory, so that $\partial_\nu J^\nu \neq 0$, and
\begin{equation}
\dfrac{d \mathcal{Q}}{dt} = \int dx \, \partial_t J^0 =  \int dx \, \partial_\nu J^\nu \neq 0 \, .
\end{equation}
We can immediately see the problem: the standard proof of the Lorentz-invariance of a charge \cite{Weinberg_book_1972} makes explicit use of the condition $\partial_\nu J^\nu =0$. In fact, one has to apply the Gauss theorem in the spacetime volume enclosed by the surfaces of constant time of Alice and Bob, which are tilted by relativity of simultaneity (see \citet{MTW_book}, figure 5.3.c). If $\partial_\nu J^\nu \neq 0$, Alice and Bob \textit{can disagree} on the average value of $\mathcal{Q}$. Our goal, now, is to quantify the disagreement.

The state $\ket{\alpha}$ models a Gaussian neutron wavepacket at rest in the origin. Following \citet{Exner:1983xu}, we assume that the wavefunction does not ``spread around'' over the decay timescale $\tau$ [i.e. $\Delta x(\tau) \approx \Delta x(0)=:\Delta x$], and we postulate (for simplicity) a purely exponential decay law. Then, working in the Heisenberg picture, we can write
\begin{equation}\label{jumb}
\bra{\alpha}J^0(t,x) \ket{\alpha} \approx \dfrac{e^{-|t|/\tau}}{\sqrt{2\pi} \, \Delta x} \exp \bigg[ -\dfrac{x^2}{2\Delta x^2}\bigg] \, ,  \spc  \bra{\alpha}J^1(t,x) \ket{\alpha} \approx 0 \, .
\end{equation}
This expression is not extremely accurate, but it ``captures the essence''. In particular, if the neutron does not decay ($\tau=+\infty$), we recover $\bra{\alpha}\partial_\nu J^\nu \ket{\alpha}=0$. But if the neutron decays (finite $\tau$), then
$
\bra{\alpha}\partial_\nu J^\nu \ket{\alpha} = \partial_t \bra{\alpha} J^0 \ket{\alpha} \neq 0 
$.
Note the presence of the absolute value in the time-dependence: equation \eqref{jumb} is valid also for negative times. To see that  $e^{-|t|/\tau}$ is the correct time-dependence for all $t\in \mathbb{R}$ (also negative), one can just invoke the Breit-Wigner formula, for a particle (at rest) with mass $M$ and decay rate $\Gamma=\tau^{-1} \,$:
\begin{equation}
\bra{\alpha}e^{iHt} \mathcal{Q}e^{-iHt}\ket{\alpha} \, \, \overset{\text{B.W.}}{\approx}\, \, \bigg| \int \dfrac{dm}{2\pi} \,  \dfrac{\Gamma e^{-imt}}{(m-M)^2 +\Gamma^2/4} \bigg|^2  = e^{-|t|/\tau} \, .
\end{equation}

Now we only have to boost from the state $\ket{\alpha}$ to the state $\ket{\Lambda\alpha}=U(\Lambda)\ket{\alpha}$. To this end, we need to remember that the current $J^\nu$ is a vector field: it transforms according to the formula \cite{srednicki_book}
\begin{equation}
U^\dagger(\Lambda) 
\begin{bmatrix}
J^0(t,x) \\
J^1(t,x) \\
\end{bmatrix}
U(\Lambda) = \gamma 
\begin{bmatrix}
 J^0(\gamma t - \gamma v x, \gamma x -\gamma v t) +  v J^1(\gamma t - \gamma v x, \gamma x -\gamma v t)  \\
 J^1(\gamma t - \gamma v x, \gamma x -\gamma v t) +  v J^0(\gamma t - \gamma v x, \gamma x -\gamma v t) \\
\end{bmatrix}  \, .
\end{equation}
Averaging the zeroth component of the above equation over $\ket{\alpha}$, we obtain
\begin{equation}
\bra{\Lambda\alpha}J^0(t,x) \ket{\Lambda\alpha} \approx \dfrac{\gamma \, e^{-\gamma| t - v x|/\tau}}{\sqrt{2\pi} \, \Delta x} \exp \bigg[ -\dfrac{\gamma^2 (x - v t)^2}{2\Delta x^2}\bigg] \, .
\end{equation}
Evaluating this formula at $t=0$, integrating over $x$, and recalling equation \eqref{QQQQQQQ}, we obtain (assume $v>0$ for clarity)
\begin{equation}\label{ghlaubing}
\bra{\Lambda\alpha}\mathcal{Q} \ket{\Lambda\alpha} \approx \int dx \, \dfrac{\gamma \, e^{-\gamma v|  x|/\tau}}{\sqrt{2\pi} \, \Delta x} \, \exp \bigg[ -\dfrac{\gamma^2 x^2 }{2\Delta x^2}\bigg] \, .
\end{equation}
Note the presence of the factor $e^{-\gamma v|  x|/\tau} \, $: by relativity of simultaneity, the time-dependence coming from $e^{-|t|/\tau}$ has been converted into a space dependence! The above integral can be solved analytically, giving
\begin{figure}
\begin{center}
\includegraphics[width=0.49\textwidth]{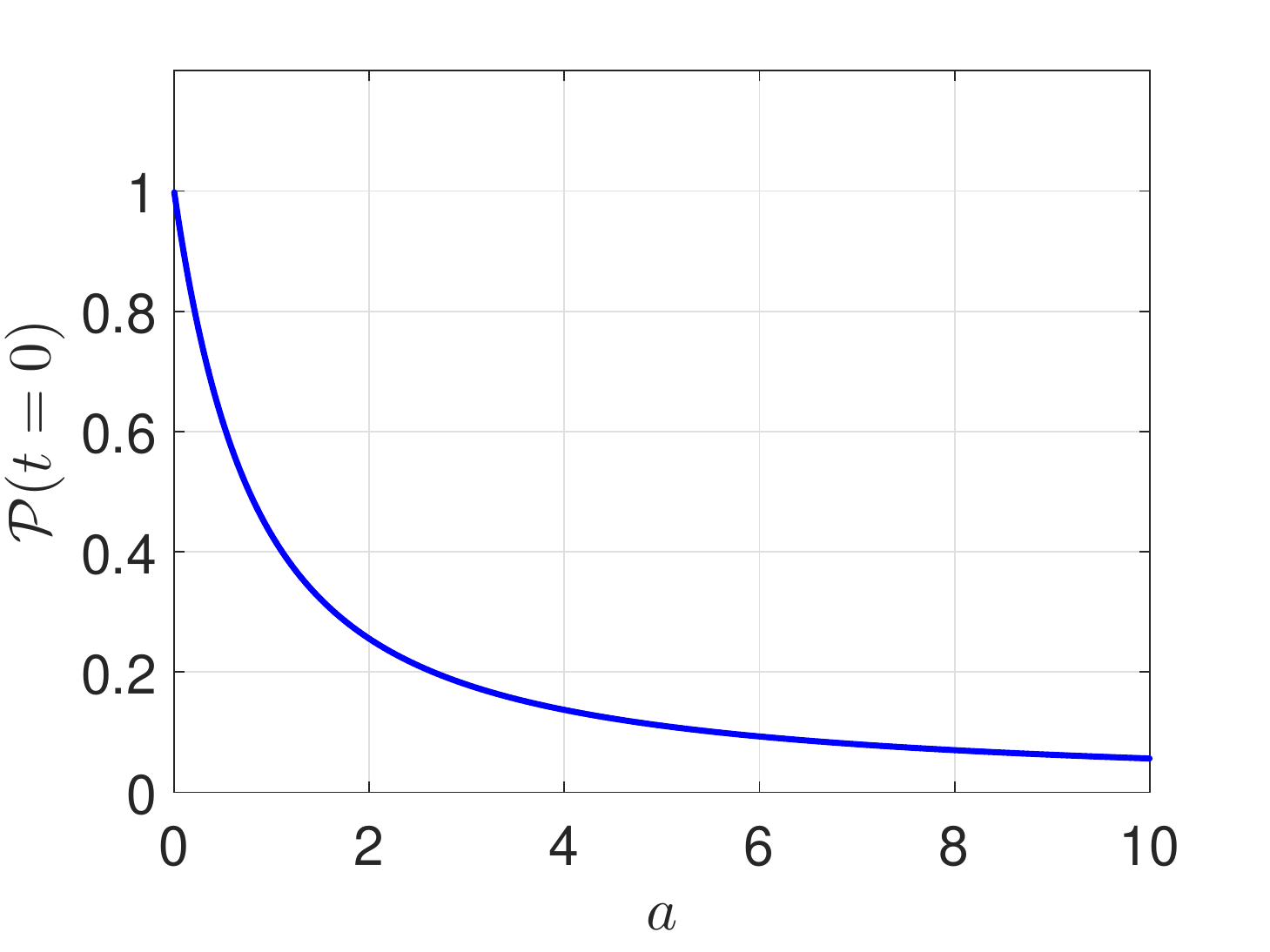}
	\caption{Graph of $\mathcal{P}(t=0)=\bra{\Lambda\alpha}\mathcal{Q} \ket{\Lambda\alpha}$ as a function of the characteristic ratio $a=v \Delta x/\tau \sqrt{2}$. The quantity $\mathcal{P}(t=0)$ expresses the probability that a boosted neutron is detected as neutron (at $t=0$). The deviations of $\bra{\Lambda\alpha}\mathcal{Q} \ket{\Lambda\alpha}$ from $1$ are possible only because the decay in time, $e^{-|t|/\tau}$, is transformed (by relativity of simultaneity) into a decay in space, $e^{-\gamma v|  x|/\tau} $. The dimensionless parameter $a$ quantifies the importance of this effect. }
	\label{fig:laP}
	\end{center}
\end{figure}
\begin{equation}\label{pooooooguabuz}
\mathcal{P}(t=0)=\bra{\Lambda\alpha}\mathcal{Q} \ket{\Lambda\alpha} \approx e^{a^2} \text{erfc} (a) \, , \spc \text{with } \, a:= \dfrac{v \, \Delta x}{\tau \sqrt{2}} \, ,
\end{equation}
where ``$\, \text{erfc} \,$'' denotes the complementary error function. We plot \eqref{pooooooguabuz} in figure \ref{fig:laP}. As we can see, if $v \rightarrow 0$, or the position uncertainty is small, or the lifetime of the particle is long, then $\bra{\Lambda\alpha}\mathcal{Q} \ket{\Lambda\alpha} \approx 1$. But if the particle is strongly delocalised and short-lived, and $v \sim 1$, the boost causes an effective decay: $\bra{\Lambda\alpha}\mathcal{Q} \ket{\Lambda\alpha} \rightarrow 0$. 

Why does $\bra{\Lambda\alpha}\mathcal{Q} \ket{\Lambda\alpha}$ depend only on $a \,$? There is a simple geometrical explanation for that. By length contraction, the uncertainty on the neutron's position is $\Delta x \, \gamma^{-1}$. On the other hand, the neutron ``is really a neutron'' only if it is found in a location where the decay factor $e^{-\gamma v|  x|/\tau} $ is close to $1\,$, otherwise (instead of detecting a neutron) we detect the decay products of the neutron: $p^++e^-+\bar{\nu}_e$. Since the decay factor $e^{-\gamma v|  x|/\tau} $ falls on a length scale $\tau(\gamma v)^{-1}$, the ratio between $\tau(\gamma v)^{-1}$ and $\Delta x \, \gamma^{-1}$ quantifies the ``neutron-ness'' of the state $\ket{\Lambda \alpha}$. Such ratio coincides with  $a^{-1}$, apart from the factor $\sqrt{2}$. Indeed, in the limit of large $a$, one has the asymptotic behaviour
\begin{equation}\label{rmgpvr}
\bra{\Lambda\alpha}\mathcal{Q} \ket{\Lambda\alpha} \approx \dfrac{1}{a \sqrt{\pi}} =   \dfrac{\tau \, \sqrt{2}}{v \, \Delta x \, \sqrt{\pi}}   \, .
\end{equation}

\section{Direct amplitude calculation}

In the previous section, we derived equation \eqref{pooooooguabuz} by expressing $\mathcal{Q}$ as the charge associated to a non-conserved current $J^\nu$. However, decay amplitudes are usually calculated in a different way. Typically, one expands the state $\ket{\alpha}$ in terms of four-momentum eigenstates, and computes the average $\bra{\alpha}e^{iHt} \mathcal{Q}e^{-iHt}\ket{\alpha}$ integrating over the four-momentum eigenbasis. It is natural to ask whether the same strategy can be adopted to compute $\bra{\alpha}U^\dagger (\Lambda) \mathcal{Q} U(\Lambda)\ket{\alpha}$. Unfortunately, this kind of calculation leads to very complicated nested integrals in momentum and mass space \cite{Stefanovich:2005ai}. Obtaining a simple and transparent formula seems out of the question. However, the qualitative features of \eqref{pooooooguabuz} remain. This is what we are going to show now. But we must warn the reader: this is a very technical and tedious calculation.

\subsection{Completeness relation}

First, let us set up some notation. We follow the same strategy of \citet{Peskin_book}, Section 7.1, equation (7.2), and we construct the completeness relation below (we work 3+1 dimensions):
\begin{equation}\label{completeness}
\mathbb{I}= \ket{\text{VAC}}\bra{\text{VAC}} + \sum_g \int \dfrac{d^3 p}{(2\pi)^3} \dfrac{\ket{ \, g,\textbf{p} \,} \bra{ \, g,\textbf{p} \,} }{2 E_g(\textbf{p})}  \, , 
\end{equation}
where $\mathbb{I}$ is the identity operator, $\ket{\text{VAC}}$ is the vacuum state, and $\ket{ \, g,\textbf{p} \,}$ are four-momentum eigenstates with mass $m_g$,
\begin{equation}\label{PJJJ}
P^j \ket{ \, g,\textbf{p} \,} = p^j \ket{ \, g,\textbf{p} \,} \spc H \ket{ \, g,\textbf{p} \,} =E_g(\textbf{p}) \ket{\, g,\textbf{p} \,} = \sqrt{m_g^2+\textbf{p}^2} \ket{\, g,\textbf{p} \,}  \, .
\end{equation}
Consistently with \citet{Peskin_book}, we had to include the denominator ``$\, 2 E_g(\textbf{p})\,$'' in \eqref{completeness}, because we are adopting the covariant normalization
\begin{equation}\label{covuzzuz}
\braket{\, \tilde{g} ,\textbf{q} \, | \, g,\textbf{p} \,}= 2 E_g(\textbf{p}) (2\pi)^3 \delta^3(\textbf{q}-\textbf{p}) \delta_{\tilde{g},g} \, .
\end{equation}
For this choice of normalization, we have the transformation law
\begin{equation}
U(\Lambda)\ket{ \, g,\textbf{p} \,}= \ket{ \, g,\Lambda\textbf{p} \,} \, , 
\end{equation}
where the notation ``$\, \Lambda\textbf{p}\,$'' just means that we construct the four-vector $(E_g,\textbf{p})$, we boost it, and we take the space part of the boosted vector. In particular, if $\Lambda$ is a boost of velocity $v$ in the $x^1$ direction, we have that
\begin{equation}\label{uber}
\Lambda \textbf{p}= (\, \gamma p^1{+}\gamma v E_g(\textbf{p}), \,  p^2, \,p^3 \,) \, .
\end{equation}

\subsection{Neutron projector}

For simplicity, we set the neutron's spin to zero, so that the projector $\mathcal{Q}$ can be expressed as follows:
\begin{equation}
\mathcal{Q} = \int \dfrac{d^3 p}{(2\pi)^3}\ket{ \, n,\textbf{p} \,} \bra{ \, n,\textbf{p} \,} \, .
\end{equation}
The states $\ket{ \, n,\textbf{p} \,}$ are not eigenstates of the Hamiltonian, because $[H,\mathcal{Q}]\neq 0$. Hence, we cannot introduce a covariant normalization, like \eqref{covuzzuz}, but we need to ``stick to the standard one'':
\begin{equation}
\braket{\, n,\textbf{q} \, | \, n,\textbf{p} \,}=  (2\pi)^3 \delta^3(\textbf{q}-\textbf{p}) \, .
\end{equation}
Note that $\ket{ \, n,\textbf{p} \,}$ are exact momentum eigenstates:
\begin{equation}\label{NPJJJ}
P^j \ket{ \, n,\textbf{p} \,} = p^j \ket{ \, n,\textbf{p} \,} \, .
\end{equation} 
This is possible because, as we said in the main text, the neutron projector must be invariant under space translations: $[\mathcal{Q},P^j]=0$. Comparing \eqref{PJJJ} and \eqref{NPJJJ}, we conclude that 
\begin{equation}\label{bridge}
\braket{\, g,\textbf{q} \, | \, n,\textbf{p} \,}=  f(g,\textbf{p}) \sqrt{2E_g(\textbf{p})} \, (2\pi)^3 \delta^3(\textbf{q}-\textbf{p}) \, .
\end{equation}
where $f(g,\textbf{p})$ is some complex distribution. Clearly, $\mathcal{Q}\ket{\text{VAC}}=0$, which implies $\braket{ \, n,\textbf{p} \, | \text{VAC}}=0$. Therefore, we can use the completeness relation to derive a condition on $\braket{\, g,\textbf{q} \, | \, n,\textbf{p} \,}$: 
\begin{equation}
(2\pi)^3 \delta^3(\textbf{q}-\textbf{p}) = \braket{\, n,\textbf{q} \, | \, n,\textbf{p} \,} = \bra{ \, n,\textbf{q} \,}\mathbb{I} \ket{ \, n,\textbf{p} \,}  = \sum_g \int \dfrac{d^3 k}{(2\pi)^3} \dfrac{\braket{\, n, \textbf{q} \,| \, g,\textbf{k} \,} \braket{ \, g,\textbf{k} \, | \,  n,\textbf{p} \,} }{2 E_g(\textbf{k})}     \, .
\end{equation}
Invoking \eqref{bridge}, this equation reduces to a normalization requirement on $f \, $:
\begin{equation}
\sum_g |f(g,\textbf{p})|^2=1  \spc \forall \, \textbf{p}\in \mathbb{R}^3 \, .
\end{equation}

\subsection{Boosted decay law}

We are finally ready to compute the formula for the decay law of a boosted neutron state. We start with a neutron state $\ket{\alpha}$, and we introduce the function
\begin{equation}\label{alpja}
\alpha(\textbf{p}):= \braket{ \, n,\textbf{p} \, | \alpha } \spc \text{with} \spc \braket{\alpha|\alpha}=\braket{\alpha|\mathcal{Q}|\alpha}= \int \dfrac{d^3 p}{(2\pi)^3} |\alpha(\textbf{p})|^2 =1 \, .
\end{equation}
Clearly, if $\ket{\alpha}$ is a neutron with probability $1$, then $\mathcal{Q}\ket{\alpha}=\ket{\alpha}$ and $\braket{\text{VAC}|\alpha}=0$. As a consequence, we can write the following chain of identities:
\begin{equation}
\ket{\alpha}= \mathbb{I}\mathcal{Q}\ket{\alpha}= \sum_g \int \dfrac{d^3 p}{(2\pi)^3}  \dfrac{d^3 q}{(2\pi)^3} \dfrac{\ket{ \, g,\textbf{p} \,}  }{2 E_g(\textbf{p})}\braket{ \, g,\textbf{p} \,| \, n,\textbf{q} \,} \braket{ \, n,\textbf{q} \, |\alpha} \, .
\end{equation}
Using \eqref{bridge} and \eqref{alpja}, this simplifies to
\begin{equation}
\ket{\alpha}= \sum_g  \int \dfrac{d^3 p}{(2\pi)^3} \, \alpha(\textbf{p}) \, f(g,\textbf{p}) \, \dfrac{\ket{ \, g,\textbf{p} \,} }{\sqrt{2E_g (\textbf{p})}} \, .
\end{equation} 
Now, we apply a Lorentz boost and we evolve the resulting state in time:
\begin{equation}\label{cambato}
e^{-iHt}U(\Lambda)\ket{\alpha} = \sum_g  \int \dfrac{d^3 p}{(2\pi)^3} \, \alpha(\textbf{p}) \, f(g,\textbf{p}) \, \dfrac{e^{-iE_g(\Lambda\textbf{p})t}\ket{ \, g,\Lambda\textbf{p} \,} }{\sqrt{2E_g (\textbf{p})}}
\end{equation}
As usual, we introduce the notation $\ket{\Lambda\alpha}:=U(\Lambda)\ket{\alpha}$. Furthermore, we change variable in the integral from $\textbf{p}$ to $\textbf{q}=\Lambda\textbf{p}$, and recall that the invariant volume element in momentum space is
\begin{equation}
\dfrac{d^3p}{2E_g(\textbf{p})}=\dfrac{d^3q}{2E_g(\textbf{q})} \, ,
\end{equation}
so that \eqref{cambato} becomes
\begin{equation}
e^{-iHt}\ket{\Lambda\alpha} = \sum_g  \int \dfrac{d^3 q}{(2\pi)^3}  \dfrac{\sqrt{2E_g(\Lambda^{-1}\textbf{q})}}{2E_g (\textbf{q})} \, \alpha(\Lambda^{-1}\textbf{q}) \, f(g,\Lambda^{-1}\textbf{q}) \, e^{-iE_g(\textbf{q})t}\ket{ \, g,\textbf{q} \,} \, . 
\end{equation}
Now we can project this state on a generic neutron state $\ket{\, n  , \textbf{k} }$. The result is
\begin{equation}\label{gringo}
\bra{\, n  , \textbf{k} } e^{-iHt}\ket{\Lambda\alpha} = \sum_g    \sqrt{\dfrac{E_g(\Lambda^{-1}\textbf{k})}{E_g (\textbf{k})}} \, \alpha(\Lambda^{-1}\textbf{k}) \, f(g,\Lambda^{-1}\textbf{k}) \, f^*(g,\textbf{k}) \, e^{-iE_g(\textbf{k})t} \, . 
\end{equation}
The decay law of the state $\ket{\Lambda\alpha}$ is the function
\begin{equation}
\mathcal{P}(t)= \bra{\Lambda\alpha} e^{iHt}\mathcal{Q}e^{-iHt}\ket{\Lambda\alpha}= \int \dfrac{d^3 k}{(2\pi)^3} |\bra{ \, n,\textbf{k} \,} e^{-iHt}\ket{\Lambda\alpha} |^2 \, .
\end{equation}
Using equation \eqref{gringo} we finally obtain an integral expression for $\mathcal{P} \, $:
\begin{equation}\label{decay}
\mathcal{P}(t)= \int \dfrac{d^3 k}{(2\pi)^3} \bigg|  \sum_g    \sqrt{\dfrac{E_g(\Lambda^{-1}\textbf{k})}{E_g (\textbf{k})}} \, \alpha(\Lambda^{-1}\textbf{k}) \, f(g,\Lambda^{-1}\textbf{k}) \, f^*(g,\textbf{k}) \, e^{-iE_g(\textbf{k})t} \bigg|^2 \, .
\end{equation}
This formula generalizes equation (13.75) of \citet{Stefanovich:2005ai}, and it is essentially exact: it is valid in any relativistic quantum theory, including QFT. The only approximation is that we have neglected the spin of the neutron. Note that the function $\alpha$ cannot be taken out of the summation, because the notation ``$\, \Lambda^{-1}\textbf{k} \,$'' stands for [see equation \eqref{uber}]
\begin{equation}
\Lambda^{-1} \textbf{k}= (\, \gamma k^1{-}\gamma v E_g(\textbf{k}), \, k^2, \, k^3 \,) \, ,
\end{equation}
and depends on $g$ through $E_g$.

\subsection{A quick check: space-translated states}

Before studying the dependence of $\mathcal{P}(0)$ on the position uncertainty, let us see what happens if we replace the state $\ket{\alpha}$ with $\ket{\beta}=e^{iP^1L}\ket{\alpha}$. First of all, we note that
\begin{equation}
\beta(\textbf{p}) = \bra{ \, n,\textbf{p} \,}e^{iP^1L}\ket{\alpha} = e^{ip^1L} \alpha(\textbf{p})  \, .
\end{equation}
Thus, in equation \eqref{decay}, we just need to replace ``$\,\alpha( \Lambda^{-1}\textbf{k}) \,$'' with
\begin{equation}
\beta(\Lambda^{-1}\textbf{k})= e^{ik^1 \gamma L} e^{-iE_g(\textbf{k})\gamma v L} \alpha(\Lambda^{-1}\textbf{k}) \, .
\end{equation} 
Now, the factor $ e^{ik^1 \gamma L}$ does not depend on $g$. Thus, in equation \eqref{decay}, we can take it out of the summation over $g$, and the absolute value cancels it. The factor $ e^{-iE_g(\textbf{k})\gamma v L} $, on the other hand, depends on $g$, and it can be combined with the exponential $e^{-iE_g(\textbf{k})t}$. The result is that the decay law of $\ket{\Lambda\beta}$ can be obtained directly from the decay law of $\ket{\Lambda\alpha}$, equation \eqref{decay}, just making the replacement
\begin{equation}
t \longrightarrow t+\gamma v L \, .
\end{equation}
This is in perfect agreement with our results of the main text: ``space translation + boost = time evolution''. From equation \eqref{decay}, we can easily see what happens if we take the limit of large $L$: the exponential  $ e^{-iE_g(\textbf{k})\gamma v L} $ becomes highly oscillating, and the contributions coming from all possible values of $g$ (which give rise to a continuum of energies) average to zero, leading to a decay.

\subsection{A couple of simplifications}

Let us go back to $\mathcal{P}(t)= \bra{\Lambda\alpha} e^{iHt}\mathcal{Q}e^{-iHt}\ket{\Lambda\alpha}$. \citet{Stefanovich:2005ai} has shown, using equation \eqref{decay}, that if $\Delta x^1 \ll \tau$ (equivalently, $a \rightarrow 0$), and the wavepacket is not too far from the origin, then $\mathcal{P}(0)\approx 1$, coherently with figure \ref{fig:laP}. We will not repeat those calculations here. We are more interested in the limit $a \rightarrow +\infty$. In this case, 
\begin{equation}\label{inequaltuz}
v \, \Delta x^1 \gg \tau \, , 
\end{equation} 
and the boost causes a decay, namely $\mathcal{P}(0)=\bra{\Lambda\alpha}\mathcal{Q}\ket{\Lambda\alpha} \approx 0$. Since we want to verify this analytically, we first need to simplify \eqref{decay}, capturing its ``essence'', without getting lost in irrelevant details.

First of all, we consider that each state $g$ has an associated rest mass $m_g$, and it is convenient to rewrite $f$, introduced in \eqref{bridge}, as a function of the mass:
\begin{equation}\label{fgtomg}
f(g,\textbf{p})=f(m_g,\textbf{p}) \, .
\end{equation} 
Of course, if the mass eigenvalues are degenerate, there may be two different states $\ket{ \, g,\textbf{p} \,}$, with same mass and momentum, but different $f$, making the above change of variables impossible. However, for our purposes, \eqref{fgtomg} is a reasonable simplification. This also allows us to convert the sum over $g$ into an integral over the masses:
\begin{equation}
\sum_g = \int dm \, \rho(m) \, .
\end{equation} 
The non-negative distribution $\rho(m)= \sum_g \delta(m-m_g)$ is the density of mass eigenstates. It is usually quite smooth close to the mass of the unstable particle, because the mass eigenstates form a continuum there \cite{Peskin_book,Stefanovich:2005ai,Exner:1983xu}. Therefore, in equation \eqref{decay}, it is convenient to build a single function out of all those contributions that do not depend on $\alpha$:
\begin{equation}
\mathcal{Z}(m,\textbf{k}):= \rho(m) \sqrt{\dfrac{E_m(\Lambda^{-1}\textbf{k})}{E_m (\textbf{k})}} \,  f(m,\Lambda^{-1}\textbf{k}) \, f^*(m,\textbf{k}) \, .
\end{equation}
Finally, it is clear that the transverse momenta $k^2$ and $k^3$ do not play an essential role in our analysis. Thus, we can just impose that there is no transverse motion:
\begin{equation}
\alpha(\textbf{p})=2\pi \, \alpha(p^1) \, \sqrt{\delta(p^2)\delta(p^3)} \, .
\end{equation}
In this way, the integrals in $dk^2$ and $dk^3$ cancel with the Dirac deltas, and we are left with an effectively one-dimensional problem. Combining these simplifications, \eqref{decay} becomes, for $t=0$,
\begin{equation}\label{pooooo}
\mathcal{P}(0) = \int \dfrac{d k}{2\pi} \, \bigg| \int dm \, \mathcal{Z}(m,k) \, \alpha\big(\gamma k -\gamma v \sqrt{m^2+k^2} \big)  \bigg|^2 \, ,
\end{equation}
where we dropped the ``$\, 1 \,$'' from ``$\, k^1 \,$'', to lighten the notation.

\subsection{Decay caused by boosts}

We need to estimate the integral in \eqref{pooooo}. We change the integration variable from $m$ to $\xi:=\gamma v \sqrt{m^2+k^2}$ (which is possible because $v \neq 0$). Then, using the relation
\begin{equation}
\dfrac{\xi^2}{\gamma^2 v^2}= m^2 +k^2  \spc \Longrightarrow \spc  \dfrac{\xi d\xi}{\gamma^2 v^2} = m dm \, ,
\end{equation}
we obtain
\begin{equation}\label{edgarpooooo}
\mathcal{P}(0) = \int \dfrac{d k}{2\pi} \, \bigg| \int \dfrac{\xi d\xi}{m\gamma^2 v^2} \, \mathcal{Z}(m,k) \, \alpha\big(\gamma k -\xi \big)  \bigg|^2 \, .
\end{equation}
Of course, in this equation, $m$ is regarded as a function of $k$ and $\xi$, through the formula
\begin{equation}\label{massinxi}
m = \sqrt{\dfrac{\xi^2}{\gamma^2 v^2} - k^2} \, .
\end{equation}
Recall that we are interested in states with very high position uncertainty: $\Delta x^1 \rightarrow +\infty$. Considering that $\Delta x^1 \Delta p^1 \sim 1$, this corresponds to taking the limit $\Delta p^1 \rightarrow 0$. Therefore, if $\ket{\alpha}$ models a neutron at rest, the function $\alpha\big(\gamma k -\xi \big) $ is well peaked around $\xi=\gamma k$, meaning that all other quantities are effectively constant over the support of $\alpha$, and can be taken out of the integral. When we do it, we must evaluate the mass $m$ at $\xi=\gamma k$, and \eqref{massinxi} becomes $m=k/ \gamma v$. Therefore, we obtain
\begin{equation}\label{lapappabuona}
\mathcal{P}(0) = \int \dfrac{d k}{2\pi v^2} \bigg| \mathcal{Z} \bigg( \dfrac{k}{\gamma v}, k \bigg) \bigg|^2  \, \times \, \bigg| \int dp  \, \alpha (p) \bigg|^2 \, .
\end{equation}
In the integral involving $\alpha$, we have performed a second change of integration variable: $p=\gamma k-\xi$ (with $dp=-d\xi$). As a result, now we have two separate integrals in \eqref{lapappabuona}, which can be easily estimated.

Let us first consider the integral in $p$. Analogously to section \ref{simpluzza}, we assume that $\ket{\alpha}$ is a Gaussian wavepacket at rest in the origin. The normalization of $\alpha(p)$ can be determined from \eqref{alpja} by comparison with the normal distribution:
\begin{equation}
\int \dfrac{dp}{2\pi} \, |\alpha(p)|^2 = 1 = \int \dfrac{dp}{\Delta p^1 \sqrt{2\pi}} \exp \bigg[ -\dfrac{p^2}{2(\Delta p^1)^2}  \bigg]  \quad \quad  \Longrightarrow \quad \quad  \alpha(p) =  \dfrac{(2\pi)^{1/4}}{\sqrt{\Delta p^1}} \exp \bigg[ -\dfrac{p^2}{4(\Delta p^1)^2}  \bigg] \, .
\end{equation}
For Gaussian wavepackets, we have that $\Delta x^1\Delta p^1 \equiv 1/2$, and the integral over $p$ in \eqref{lapappabuona} becomes
\begin{equation}\label{adp}
\int dp \, \alpha(p) = 2^{5/4} \pi^{3/4} \sqrt{\Delta p^1} = \dfrac{(2\pi)^{3/4}}{\sqrt{\Delta x^1}} \, .
\end{equation}
Let us now focus on the first integral in \eqref{lapappabuona}. It is well known \cite{Stefanovich:2005ai} that the absolute value of $f(m,\textbf{p})$ has a very weak dependence on $\textbf{p}$. Actually, in some interaction models, one can even show that $|f(m,\textbf{p})|$ is a pure function of $m$. Therefore, in our estimates, we can just replace $|\rho(m)   f(m,\Lambda^{-1}\textbf{k}) \, f^*(m,\textbf{k})|$ with $\rho(m) \, | f(m,0)|^2$. On the other hand, the function $\rho(m) \,|f(m,0)|^2$ is just the distribution of energy of the neutron at rest (equivalently, its distribution of mass). To see this, one can average an arbitrary power of the Hamoltonian over $\ket{ \, n,0 \,} \,$:
\begin{equation}
\dfrac{\bra{ \, n,0 \,} H^N \ket{ \, n,0 \,}}{\braket{ \, n,0 \, | \, n,0 \,}} =   \sum_g \int \dfrac{d^3 p \, E_g^N(\textbf{p})}{(2\pi)^6 \delta^3(0)} \dfrac{|\braket{ \, g,\textbf{p} \, | \, n,0 \,} |^2}{2 E_g(\textbf{p})} = \int dm \, m^N \, \rho(m) \, |f(m,0)|^2  \, .
\end{equation} 
Therefore, the qualitative behaviour of $|\mathcal{Z}(m,\textbf{k})|^2$ is very well capture by the Breit-Wigner approximation:
\begin{equation}
|\mathcal{Z}(m,\textbf{k})|^2 \approx \dfrac{E_m(\Lambda^{-1}\textbf{k})}{E_m (\textbf{k})} \bigg| \dfrac{\Gamma/2\pi}{(m-M)^2 +\Gamma^2/4} \bigg|^2  \, ,
\end{equation}
where $M$ is the neutron's average mass, and $\Gamma=\tau^{-1}$ its decay rate. 
Furthermore, note that, since in \eqref{lapappabuona} we evaluate $\mathcal{Z}$ for $k^2=k^3=0$ and $m=k/\gamma v$, the ratio $E_m(\Lambda^{-1}\textbf{k})/E_m (\textbf{k})$ is equal to $\gamma^{-1}$. This enables us to perform the integration analytically:
\begin{equation}\label{caputanuncino}
\int \dfrac{d k}{2\pi v^2} \bigg| \mathcal{Z} \bigg( \dfrac{k}{\gamma v}, k \bigg) \bigg|^2 = \int \dfrac{d k}{2\pi \gamma v^2} \, \bigg| \dfrac{\Gamma/2\pi}{(k/\gamma v -M)^2 +\Gamma^2/4} \bigg|^2 = \dfrac{\tau}{2\pi^2 v} \, .
\end{equation}
Plugging \eqref{adp} and \eqref{caputanuncino} into \eqref{lapappabuona}, we finally obtain
\begin{equation}\label{pnubo}
\mathcal{P}(0) \overset{a \rightarrow \infty}{\approx} \dfrac{\tau \sqrt{2}}{ v \, \Delta x^1 \, \sqrt{\pi}} \, .
\end{equation}
We have recovered equation \eqref{rmgpvr}. It is quite surprising that two such different approaches resulted in exactly the same final formula, with the factors $\sqrt{2}$ and $\sqrt{\pi}$ in the right position. This is the magic of Quantum Field Theory!

\end{document}